\documentclass[letterpaper,twocolumn,10pt]{article}
\usepackage{usenix}

\newif\ifsubmission
\submissiontrue

\usepackage{graphicx}
\usepackage[inline]{enumitem}
\usepackage{lstautogobble}
\usepackage{listings}
\usepackage{ebproof}
\usepackage[acronym]{glossaries}
\glsdisablehyper
\usepackage{hyperref}
\hypersetup{
  colorlinks,
  linkcolor={black!65}
}
\usepackage{xurl}
\usepackage{algorithmic}
\usepackage{algorithm}
\usepackage{mathtools}
\usepackage{amsthm}
\usepackage{subcaption}
\usepackage{xspace}
\usepackage{amssymb}
\usepackage{booktabs}
\usepackage{tabularx}
\usepackage{pifont}
\usepackage{multirow}

\usepackage[capitalize,nameinlink]{cleveref}
\usepackage{proof}
\usepackage{amsthm}
\usepackage[dvipsnames]{xcolor}
\ifsubmission
  \newcommand{\comment}[3]{}
\else
  \newcommand{\comment}[3]{\textcolor{#2}{[#1: #3]}}
\fi

\newcommand{\felix}[1]{\comment{Felix}{OliveGreen}{#1}}
\newcommand{\david}[1]{\comment{David}{NavyBlue}{#1}}
\newcommand{\codex}[1]{\comment{Codex}{Purple}{#1}}

\crefname{ALC@unique}{line}{Lines}
\Crefname{ALC@unique}{Line}{Lines}
\crefname{point}{point}{points}
\Crefname{point}{Point}{Points}

\newacronym{DoS}{DoS}{Denial-of-Service}
\newacronym{DNS}{DNS}{Domain Name System}
\newacronym{RR}{RR}{resource record}
\newacronym{TLD}{TLD}{top-level domain}
\newacronym{DNSSEC}{DNSSEC}{Domain Name System Security Extensions}
\newacronym{MGU}{MGU}{most general unifier}
\newacronym{CLP}{CLP}{constraint logic program}
\newacronym{IETF}{IETF}{Internet Engineering Task Force}
\newacronym[longplural={Requests for Comments}]{RFC}{RFC}{Request for Comments}

\newcommand{\varstyle}[1]{\ensuremath{\mathsf{#1}}\xspace}
\newcommand{\funcstyle}[1]{\ensuremath{\mathsf{#1}}}
\newcommand{\choices}[1]{\ensuremath{\mathit{#1}}}

\newcommand{\vars}{\ensuremath{\mathcal{V}}}
\newcommand{\varsf}{\ensuremath{\funcstyle{vars}}}
\newcommand{\consts}{\ensuremath{\mathcal{C}}}
\newcommand{\funcs}{\ensuremath{\mathcal{G}}}
\newcommand{\preds}{\ensuremath{\mathcal{P}}}
\newcommand{\cons}{\ensuremath{\cdot}}

\newcommand{\emptylist}{\ensuremath{[]}}
\newcommand{\ts}{\ensuremath{\mathcal{T}}}
\newcommand{\formulas}{\ensuremath{\mathcal{F}}}
\newcommand{\lrule}{\ensuremath{\leftarrow}}
\newcommand{\symdiff}{\ensuremath{\mathbin{\Delta}}}
\newcommand{\disjcup}{\ensuremath{\mathbin{\dot\cup}}}
\newcommand{\pto}{\ensuremath{\rightharpoonup}}
\newcommand{\dom}{\ensuremath{\funcstyle{dom}}}
\newcommand{\img}{\ensuremath{\funcstyle{img}}}

\newcommand{\C}{\ensuremath{\mathfrak{C}}}
\newcommand{\lfp}{\ensuremath{\funcstyle{lfp}}}

\ExplSyntaxOn
\seq_new:N \l__lprule_lines_seq

\cs_new_protected:Npn \lprule_typeset:n #1
  {
    \seq_set_split:Nnn \l__lprule_lines_seq { \\ } { #1 }
    \begin{array}{l}
      \seq_map_indexed_inline:Nn \l__lprule_lines_seq
        {
          \int_compare:nNnTF { \seq_count:N \l__lprule_lines_seq } = { 1 }
            { \tl_trim_spaces:n { ##2 }. }
            {
              \int_compare:nNnTF { ##1 } = { 1 }
                { \tl_trim_spaces:n { ##2 } \lrule \\ }
                {
                  \quad \tl_trim_spaces:n { ##2 }
                  \int_compare:nNnTF { ##1 } = { \seq_count:N \l__lprule_lines_seq }
                    { . }
                    { , \\ }
                }
            }
        }
    \end{array}
  }

\cs_new_protected:Npn \lprule_typeset_display:n #1
  {
    \mode_if_math:TF
      {
        \mathmakebox[\dimexpr\displaywidth-2\multlinegap\relax][c]
          { \lprule_typeset:n { #1 } }
      }
      {
        \[
          \mathmakebox[\dimexpr\displaywidth-2\multlinegap\relax][c]
            { \lprule_typeset:n { #1 } }
        \]
      }
  }

\NewDocumentEnvironment{lprule}{+b}
  { \lprule_typeset_display:n { #1 } }
  { }
\ExplSyntaxOff

\newcommand{\labels}{\ensuremath{\mathcal{L}}}
\newcommand{\names}{\ensuremath{\mathcal{N}}}
\newcommand{\data}{\ensuremath{\mathcal{D}}}

\newcommand{\types}{\ensuremath{\mathbb{T}}}
\newcommand{\typesSec}{\ensuremath{\mathbb{T}_\funcstyle{SEC}}}
\newcommand{\mathdot}{\ensuremath{\text{.}}}
\newcommand{\wc}{\ensuremath{\ast}}
\newcommand{\name}[1]{\ensuremath{\texttt{#1}}}

\newcommand{\tdots}{\text{\dots}}

\newcommand{\NS}{\varstyle{NS}}
\newcommand{\A}{\varstyle{A}}
\newcommand{\AAAA}{\varstyle{AAAA}}
\newcommand{\TXT}{\varstyle{TXT}}
\newcommand{\CNAME}{\varstyle{CNAME}}
\newcommand{\DNAME}{\varstyle{DNAME}}
\newcommand{\RRSIG}{\varstyle{RRSIG}}
\newcommand{\DNSKEY}{\varstyle{DNSKEY}}
\newcommand{\DS}{\varstyle{DS}}
\newcommand{\MX}{\varstyle{MX}}
\newcommand{\SOA}{\varstyle{SOA}}
\newcommand{\NSEC}{\varstyle{NSEC}}
\newcommand{\NSECIII}{\varstyle{NSEC3}}
\newcommand{\NSECIIIParam}{\varstyle{NSEC3PARAM}}

\newcommand{\NXDOMAIN}{\varstyle{NXDOMAIN}}

\newcommand{\SERVFAIL}{\varstyle{SERVFAIL}}

\newcommand{\namerep}{\ensuremath{\mathsf{name}}}
\newcommand{\datarep}{\ensuremath{\mathsf{data}}}

\newcommand{\wcc}{\ensuremath{\funcstyle{C}}}

\newcommand{\resolvable}{\ensuremath{\funcstyle{resolvable}}}
\newcommand{\resolves}{\ensuremath{\funcstyle{resolves}}}

\newcommand{\referral}{\ensuremath{\funcstyle{referral}}}

\newcommand{\digestValid}{\ensuremath{\funcstyle{digestValid}}}
\newcommand{\signatureValid}{\ensuremath{\funcstyle{signatureValid}}}
\newcommand{\nsecUnsigned}{\ensuremath{\funcstyle{nsecUnsigned}}}
\newcommand{\closestEncloser}{\ensuremath{\funcstyle{closestEncloser}}}
\newcommand{\nsecIIIUnsigned}{\ensuremath{\funcstyle{nsec3Unsigned}}}
\newcommand{\namec}{\ensuremath{\funcstyle{isName}}}
\newcommand{\member}{\ensuremath{\funcstyle{member}}}

\newcommand{\solve}{\ensuremath{\funcstyle{solve}}}

\newcommand{\MGU}{\ensuremath{\funcstyle{MGU}}}

\newtheorem{thm}{Theorem}
\newtheorem{lem}{Lemma}

\newcommand{\cmark}{\ding{51}}
\newcommand{\xmark}{\ding{55}}

\theoremstyle{definition}
\newtheorem{definition}{Definition}[section]

\begin{document}

\date{}

\title{Declarative DNS -- Resolving Recursive Resolution}

\author{
{\rm Felix Linker}\\
ETH Zurich
\and
{\rm Yue Huang}\\
HKUST(GZ)
\and
{\rm David Basin}\\
ETH Zurich
\and
{\rm Huayi Duan}\\
HKUST(GZ)
} 

\maketitle

\begin{abstract}
DNS zone configuration errors are a frequent source of outages and frustration, particularly in combination with DNSSEC.
Research has attempted to address the problem of zone configuration errors using formal methods, but existing proposals
\begin{enumerate*}[label=(\roman*)]
    \item only verify the absence of specific, known configuration errors, and
    \item do not consider DNSSEC.
\end{enumerate*}

We present a new formal model of the DNS that covers most of DNSSEC, making it one of the most complete DNS models to date.
In addition, we present a new tool, DecDNS, that performs a \emph{resolvability change-impact analysis} of zone configuration updates using our model.
We show that such a change-impact analysis can identify configuration errors without explicitly searching for them.

We evaluate DecDNS both formally and empirically.
We prove that DecDNS is sound and complete and show that its results agree with widely used DNS software.
Finally, we explain how our approach can be integrated into operational practice, helping guide zone management in the real world.
\end{abstract}

\paragraph{Availability}
Our tool, DecDNS, is available on GitHub:
\begin{center}
    \url{https://github.com/felixlinker/decdns}
\end{center}

\section{Introduction}
The \gls{DNS} is a fundamental Internet technology.
So fundamental that the unavailability of the \gls{DNS} can be practically indistinguishable from the unavailability of the Internet.
For example, some governments censor the Internet solely by filtering the \gls{DNS} \cite{AFNICFiltering,SSACFiltering,DNSCensorship1,DNSCensorship2}.
Although ``censored'' services are technically still reachable, many clients and users cannot reach them anymore.

At first glance, the \gls{DNS} appears simple.
Data is associated with \emph{names} such as \name{example.com}.
All names form a tree, and sets of connected nodes form so-called \emph{zones}.
Different zones are served by different \emph{name servers}, and \emph{recursive resolvers} query name servers to obtain the data associated with names.
Underlying this simple concept, however, is considerable complexity.
The \gls{DNS}'s two original standards were published in 1987 and, since then, it has been extended numerous times.
As of fall 2026, the \gls{IETF}, which maintains the \gls{DNS}, has published 298 documents, called \glspl{RFC}, on the \gls{DNS} \cite{DNSRFCs}.

One of the most substantial extensions to the \gls{DNS} are the \gls{DNSSEC}.
When using \gls{DNSSEC}, name servers sign the data they serve and publish public key material for verification.
\Gls{DNSSEC} drastically increases the complexity of the \gls{DNS} and thus the likelihood of configuration errors.
\felix{There is work on the prevalence of DNSSEC configuration errors, which we could cite here. For example, \cite{DNSSECEmpirial,DNSSECEmpirical2}.}

It thus comes as no surprise that, given the complexity of the \gls{DNS}, configuration errors are common and they can have severe consequences.
For example, in May 2026, roughly 2/3 of the \name{.de} \gls{TLD} became unavailable to many resolvers \cite{DenicOutage} due to a \gls{DNSSEC} configuration error.
The \name{.de} \gls{TLD} is one of the most frequently queried \glspl{TLD} \cite{DenicOutageCloudflare}, and a configuration error in the published public key material used to authenticate data in the \name{.de} zone led to a 3-hour outage.

There are proposals for detecting configuration errors.
For example, some tools test a domain name for certain errors \cite{DNSViz,Zonemaster}.
There is also a large body of work using formal reasoning to detect configuration errors \cite{DNSMaude,DelphiPaper,GRoot,VeriDNS,DNSOctopus}.
Such formal approaches generally promise to provide ``complete coverage,'' i.e., modulo abstractions and assumptions of the underlying model and theory, they can provide mathematical certainty.
However, existing formal approaches all share two limitations.
\begin{enumerate*}[label=(\roman*)]
  \item \label[point]{itm:absence-verification} They suggest verifying the absence of particular, concrete configuration errors.
  \item They do not consider \gls{DNSSEC}, one of the largest extensions to the \gls{DNS}.
\end{enumerate*}

Note that \cref{itm:absence-verification} above is in stark contrast to formal methods' promise of complete coverage.
Since existing approaches only verify the absence of particular classes of configuration errors, they do not capture configuration errors generally and miss every configuration error not encoded directly in the model.

In this work, we address the above two limitations.
We present a new, declarative formal model and a tool called DecDNS (declarative \gls{DNS}) to reason about \gls{DNS} configuration errors.
Using DecDNS, one can identify configuration errors without explicitly encoding them in the model.
Moreover, our model and DecDNS provide support for \gls{DNSSEC}.

\paragraph{Resolvability Change-Impact Analysis}
In our approach, we take a new perspective on \gls{DNS} configurations and consider the property of \emph{resolvability}.
We argue that a zone configuration can be characterized by which queries should have a positive response and are thus resolvable.
If such a characterization were given, one could detect configuration errors by computing all queries that have a positive response.
Any difference to the set of queries that should have a positive response would point to a configuration error.

Such a characterization, though, is hard to obtain.
Having the user state all queries that should be resolvable is infeasible and likely a task just as error-prone as configuring the zone itself.
We therefore use the old configuration as a baseline and perform a \emph{resolvability change-impact analysis} on a zone configuration update.
Given two zone configurations, new and old, a change-impact analysis computes which queries have a positive response in exactly one of the two configurations.
Users can then inspect the result of this analysis, compare it with their expectations, and consequently identify configuration errors.

A change-impact analysis, though, will not be able to identify \emph{all} configuration errors.
For example, resolvability disregards whether the data associated with a name is correct, e.g., whether the IP address associated with \name{example.com} indeed routes to \name{example.com}.
Moreover, we only detect configuration errors when they are introduced but not when an update preserves them.
Nevertheless, a resolvability change-impact analysis is a beneficial tool as it characterizes a configuration update and is not limited to any particular sort of configuration error.

Additionally, a change-impact analysis is particularly useful for configuration changes that should not impact resolvability, i.e., after which the set of resolvable queries should remain unchanged.
Given such a change, a non-empty change-impact analysis immediately highlights a configuration error.
Examples of such changes include changing which name servers serve particular zones, changing how the \gls{DNS} is divided into zones, or changing a zone's verification keys, which is the kind of change that led to the \name{.de} \gls{TLD} outage.

\paragraph{Overview of Our Approach}
Given a zone configuration, DecDNS compiles it to a \gls{CLP}.
This declarative specification captures which queries have a positive response in that zone configuration.
We abstract from details of the underlying transport mechanisms, e.g., packet sizes and headers, and instead model what a recursive resolver would conclude in response to arbitrary queries if it were given the input zone configuration.
We capture, though, that recursive resolvers must first discover name servers and obtain their network addresses before they can send them queries.
Given two such \glspl{CLP}, one corresponding to an old zone configuration and the other to a new version of the configuration, DecDNS computes a change-impact analysis.


Modeling the \gls{DNS} and recursive resolution as a \gls{CLP} has many benefits.
Whereas standard logic programs can only use logical predicates, \glspl{CLP} can additionally use constraints, which are interpreted by a constraint solver.
We use constraints to \begin{enumerate*}[label=(\roman*)]
    \item efficiently represent and reason about very large zones symbolically, and
    \item model cryptographic checks like signature verification.
\end{enumerate*}
Symbolic representations of large sets and cryptographic checks would be cumbersome to model in ordinary logic programs.

Our formal model only uses standard \gls{CLP} features and is interoperable with existing logic programming software, for example SWI-Prolog \cite{SWIProlog}, for which we provide an integration.
Interoperability with existing software facilitated development initially, but also is independently desirable as it supports the integration of our model or parts thereof into other applications.
Moreover, to prove the soundness and completeness of our change-impact analysis algorithm, we can leverage existing soundness and completeness results for \glspl{CLP} \cite{CLP}.

\paragraph{Contributions}
We present the following contributions.
\begin{itemize}
    \item We formally model the \gls{DNS} and many of its extensions as a \gls{CLP}, providing one of the most detailed formal models of the \gls{DNS} to date.
    In particular, our model covers the \gls{DNSSEC} standards.

    Our model is an executable, declarative specification of the \gls{DNS}, which thus promises to be relatively easy to reason about.
    It could serve as a reference implementation and a basis for further experiments with extensions or variants of the \gls{DNS}.

    \item We present an algorithm to compute a change-impact analysis from two \glspl{CLP} that represent the new and old zone configurations.
    We provide soundness and completeness results for our algorithm with respect to the well-established semantics of \glspl{CLP} \cite{CLP}.

    \item We show that our approach is practical by implementing DecDNS, which can compile a \gls{CLP} from a zone configuration, following our formal model, and reason about such \glspl{CLP}.
    DecDNS provides a mode that computes a change-impact analysis and a querying mode that resolves \gls{DNS} queries using live \gls{DNS} infrastructure.
    The querying mode, in particular, shows that our formal model is not only quite detailed, but that it can also interoperate with live \gls{DNS} infrastructure.
    It can be used, relatively straightforwardly, though somewhat inefficiently, to resolve \gls{DNS} queries on real systems.

    \item Although we prove the soundness and completeness of our algorithm, no theoretical analysis can establish whether our formal model is faithful to the \gls{DNS} standards.
    We thus also empirically evaluate DecDNS and show that its results agree with those of major \gls{DNS} software.
    First, we evaluate whether what DecDNS claims to be resolvable is indeed resolvable by real recursive resolver software.
    Second, we evaluate whether DecDNS's querying mode computes the same responses to queries as other resolvers.
    We also provide a performance evaluation of DecDNS.

    \item Finally, we show how to use DecDNS to detect configuration errors.
    We show which kinds of errors it can detect and under what circumstances.
\end{itemize}

\paragraph{Structure}
We begin with a brief example of how DecDNS works and can be used in \cref{sec:example}.
Then, we provide relevant background on the syntax of \glspl{CLP} and the \gls{DNS} in \cref{sec:bg}.
In \cref{sec:model}, we present how we compile a zone configuration to a \gls{CLP}.
Then, in \cref{sec:reasoning}, we present our algorithm to compute a change-impact analysis of two \glspl{CLP} representing the old and new zone configurations.
We present soundness and completeness results as well as relevant background on \gls{CLP} semantics in \cref{sec:soundness}.
We present our implementation in \cref{sec:implementation}.
In \cref{sec:evaluation}, we evaluate DecDNS's conformance with the \gls{DNS} and show which configuration errors it can detect.
We discuss related work in \cref{sec:rw} and draw conclusions in \cref{sec:conclusion}.

\section{An Example of DecDNS in Action}
\label{sec:example}
We start by providing a brief example of how DecDNS works, how it can be used, and how it can detect potential configuration errors.
Consider the following zone, in which the label \name{lo[{\dots}]ong} has 63 characters:

\begin{verbatim}
*.short. TXT "hello"
*.lo[...]ong. TXT "hello"
\end{verbatim}

This zone associates two so-called \emph{wildcard owner names}, those beginning with \name{*}, with \TXT{} records, i.e., the textual data ``hello.''
The semantics of wildcard owner names are complex, and we will explain them in further detail in \cref{sec:dns-bg}, but put simply, queries for \TXT{} records to a subdomain of \name{short.} or \name{lo[{\dots}]ong.} can be answered with ``hello'' in this configuration.

Now suppose the zone is updated as follows.
\begin{verbatim}
short. DNAME lo[...]ong.
*.lo[...]ong. TXT "hello"
\end{verbatim}

In this update, one wildcard \TXT{} record was replaced with a \DNAME{} alias to \name{lo[...]ong.}.
\DNAME{} aliases map each subdomain of an aliased name, here \name{short.}, to the matching subdomain of the alias target.
This \DNAME{} record, for example, would map the query \name{a.b.c.short.} to \name{a.b.c.lo[...]ong.}.
Thus, intuition suggests that this update should preserve the resolvability of subdomains of \name{short.} since the new alias target has the same record associated with it as the previous wildcard record.

However, one must also take into consideration that \gls{DNS} names have a maximum length, and DecDNS would highlight exactly this point in its change-impact analysis for the above two zones.
Concretely, DecDNS would return:
\begin{lprule}
    \resolvable(\TXT{}, [ \name{short}, L \mid T ], \bot) \\
    L \neq \wc \\
    \neg \namec([ \name{lo[...]ong}, L \mid T ]) \\
    \namec([\name{short}, L \mid T]).
\end{lprule}

Above is a formula which models when queries for \TXT{} records to any subdomain of \name{short.} are resolvable in the original configuration but not after the update.
The term $[ \name{short}, L \mid T ]$ represents an arbitrary subdomain of \name{short.}, and is a list of labels.
Concretely, $L$ is a variable representing the label directly below \name{short.}, and $T$ is an arbitrary list suffix, i.e., represents further labels below $L$\name{.short.}.
The above rule expresses that \TXT{} queries to subdomains of \name{short.} are not resolvable whenever
\begin{enumerate*}[label=(\alph*)]
    \item that subdomain is not at or below \name{*.short.} ($L \neq \wc$),
    \item the mapped query is too long, i.e., the respective subdomain of \name{lo[...]ong.} is not a name, and
    \item the initial query name is of valid length, i.e., the respective subdomain of \name{short.} is a name.
\end{enumerate*}

The administrator of the respective zone could examine this output and determine whether this loss in resolvability is acceptable and matches their expectations of the update.

While the above example is conceptually simple and small, it highlights how even such small zones can exhibit subtle behavior that can be easy to miss.
It also shows the unique benefits of a change-impact analysis.
The loss in resolvability in this example, as characterized above, would be hard to formalize in an explicit property that checks for particular configuration errors.
A change-impact analysis can highlight the subtle interactions that different parts of the \gls{DNS} standards have and help zone administrators understand the consequences of their actions.

\section{Background}
\label{sec:bg}
\subsection{Mathematical Notation}

\subsubsection{Functions and Relations}

We use $\pto$ to denote a partial function, i.e., a function that may be undefined for parts of its domain.
Note that every total function is also a partial function.
We will sometimes define partial functions explicitly as a set of mappings, e.g., $f = \{ 1 \mapsto 2 \}$ is a partial function $\mathbb{N} \pto \mathbb{N}$ such that $f(1) = 2$ and $f(x)$ is undefined for all $x \neq 1$.

Following functional programming conventions, we only use the right-associative symbols $\to$ and $\pto$ to define functions, even when they have multiple arguments.
For example, $f : A \to B \to C$ is a function with two arguments, mapping to $C$.
This notation permits \emph{partial application}, e.g., $f(a)$ is the function $f': B \to C$ s.t. $f'(b) = f(a, b)$.
We will sometimes treat functions like relations.
For example, if $f(a, b) = c$, then $(a,b,c) \in f$.

Let $f : A \pto B$ be a partial function.
We write $\dom(f)$ for $f$'s domain, i.e., $\dom(f) \subseteq A$ and $a \in \dom(f)$ if $f(a)$ is defined.
We write $\img(f)$ for $f$'s image, i.e., $\img(f) \subseteq B$ and $b \in \img(f)$ if there exists $a \in \dom(f)$ such that $f(a) = b$.
We write $f|_X$ to denote the restriction of a partial function $f$'s domain to $X$.
That is: $f|_X(x) = f(x)$ whenever $ x \in X$ and undefined otherwise.

\subsubsection{Terms and Unification}

We follow \cite{LogicProgramming} in the definitions below.
Let $\funcs$ be a set of function symbols.
We write $s/n$ to denote that a symbol $s$ has arity $n$.
Given a set of function symbols $\funcs$, variables $\vars$, and constants $\consts$, we define the set of \emph{terms} $\ts$ over those sets inductively as follows:
\[
  \begin{array}{ll}
    v \in \ts & v \in \vars \\
    c \in \ts & c \in \consts \\
    f(t_1, \dots, t_n) \in \ts & f/n \in \funcs, t_1, \dots, t_n \in \ts \\
  \end{array}
\]

We use $\varsf(t)$ to denote the set of variables occurring in a term $t$.
A term $t$ is \emph{ground} if $\varsf(t) = \emptyset$.
We use uppercase letters for variables in the logic program and lowercase letters for metavariables in our definitions.
For example, $X$ is a variable term and $v$ is a variable.
We will only refer to constants when they have been explicitly introduced as such.
Sometimes, the sets of function symbols, variables, and constants will be part of a signature $\Sigma$, in which case we write $\ts_\Sigma$ for all terms over that signature.

A \emph{substitution} is a partial function $\theta : \vars \pto \ts$, which maps variables to terms.
We homomorphically extend substitutions to also apply to sets of terms.
A substitution is a \emph{valuation} if it is total and every term in its image is ground.
For two substitutions $\theta_1, \theta_2$, we write $\theta_1\theta_2$ to denote their \emph{composition}, defined as:
\[
    \theta_1\theta_2 = \theta_2|_{\dom(\theta_2) \setminus \dom(\theta_1)} \cup (\theta_2 \circ \theta_1)
\]

For example, for two constants $a$ and $b$:
\begin{gather*}
    \theta_1 = \{ X \mapsto f(Y), Y \mapsto Z \} \quad \theta_2 = \{ X \mapsto a, Y \mapsto b, Z \mapsto Y \} \\
    \theta_1\theta_2 = \{ X \mapsto f(b), Y \mapsto Y, Z \mapsto Y \}
\end{gather*}

For a set of terms $T$, a substitution $\theta$ is a \emph{unifier} if $\theta(T)$ is a singleton set, i.e., all pairs of terms in $T$ are equal after applying $\theta$.
A unifier $\theta$ is the \emph{most general unifier} for $T$ if for all unifiers $\theta'$ there exists a substitution $\sigma$ such that $\theta' = \theta\sigma$.
It is well-known that for any two terms $t_1$ and $t_2$, a most general unifier exists if and only if $t_1$ and $t_2$ unify \cite{LogicProgramming}, which we denote by $\MGU(t_1, t_2)$.

\subsection{Constraint Logic Programs}
\label{sec:bg-lp}

We will model the \gls{DNS} using \acrfullpl{CLP}.
Logic programs are sets of Horn clauses, which in turn are first-order logic formulas of the form $\forall \vec{x}. b_1 \land \dots \land b_n \implies h$.
In a Horn clause, every variable is universally quantified.
Moreover, a Horn clause has a single predicate $h$ on the implication's right-hand side, the \emph{head}, and a finite (possibly empty) set of predicates on the implication's left-hand side, the \emph{body}.
In this section, we will introduce the syntax of \glspl{CLP} following \cite{CLP}.

\subsubsection{Logic Programs}

Logic programs are defined with respect to a signature $\Sigma = (\vars, \consts, \funcs, \preds)$ for a set of variables $\vars$, constants $\consts$, function symbols $\funcs$, as before, and a set of predicate symbols $\preds$.
The set of formulas $\formulas_\Sigma$ over $\Sigma$ is inductively defined as:
\[
  \begin{array}{ll}
    p(t_1, \dots, t_n) \in \formulas_\Sigma & p/n \in \preds, t_1, \dots, t_n \in \ts_\Sigma \\
    \phi_1 \lor \phi_2 \in \formulas_\Sigma & \phi_1, \phi_2 \in \formulas_\Sigma \\
    \phi_1 \land \phi_2 \in \formulas_\Sigma & \phi_1, \phi_2 \in \formulas_\Sigma \\
    \neg \phi \in \formulas_\Sigma & \phi \in \formulas_\Sigma \\
    \forall v. \phi \in \formulas_\Sigma & v \in \vars, \phi \in \formulas_\Sigma \\
    \exists v. \phi \in \formulas_\Sigma & v \in \vars, \phi \in \formulas_\Sigma
  \end{array}
\]

A formula is an \emph{atom} if it has a predicate as a top-level symbol.
For a formula $\varphi$, we overload $\varsf(\varphi)$ to denote all free variables occurring in $\varphi$.
A formula $\varphi$ is \emph{closed} if $\varsf(\varphi) = \emptyset$.
A formula is a \emph{Horn clause} if it is closed and for $0 \leq n, 1 \leq m$ and atoms $\phi_1, \dots, \phi_m$ it is of the form:
\[
  \forall v_1. \dots \forall v_n. \phi_1 \lor \neg \phi_2 \lor \dots \lor \neg \phi_m.
\]

Following standard logic programming conventions, we will write Horn clauses as inference rules.
The above Horn clause can be written equivalently as:
\[
  \phi_1 \lrule \phi_2, \dots, \phi_m.
\]

If $m = 1$, we omit $\lrule$ and simply write $\phi_1$.
We call such rules \emph{facts}.
Note that free variables are implicitly universally quantified.
A \emph{logic program} is a set of Horn clauses $\Gamma \subseteq \formulas_\Sigma$.

We will treat sequences of conjuncts and sets of conjuncts interchangeably and use $\top$ to denote the empty set of conjuncts.
For example:
\begin{gather*}
    \begin{array}{ccc}
        \top = \emptyset, &
        \phi \lrule \top = \phi,
    \end{array} \\
    \phi_1 \lrule \phi_2, \dots, \phi_n = \phi_1 \lrule \{ \phi_2, \dots, \phi_n \}.
\end{gather*}

\paragraph{Notation}
To limit the naming of variables, we will write $\_$ to denote a fresh, universally quantified variable that occurs nowhere else.
For example, $\forall x. \forall y. p(x, y)$ and $p(\_, \_)$ are equal up to renaming of quantified variables.
In this paper, we will use the function symbols $\cons/2, \emptylist/0, 1/0, 0/0$ where $\cons$ prepends an element to a list, $\emptylist$ denotes the empty list, $1$ denotes the bit 1, and $0$ denotes the bit 0.
Following Prolog syntax, $[x_1, \dots, x_n]$ denotes a list with $n$ elements and $[ x_1, \dots, x_n \mid y ]$ denotes a list with tail $y$.
This means:
\begin{gather*}
    [x_1, \dots, x_n] = x_1 \cons (x_2 \cons (\dots \cons (x_n \cons \emptylist))) \\
    [ x_1, \dots, x_n \mid y ] = x_1 \cons (x_2 \cons (\dots \cons (x_n \cons y))).
\end{gather*}

\subsubsection{Resolution}
\label{sec:logic-resolution}

Logic programs are typically used by querying them with a \emph{goal}.
A goal is an atom potentially containing free variables, and querying the logic program should determine whether the goal is entailed by the program.
For example, consider the following program $P$, which formalizes when a Peano number is even and when all elements of a list are even:
\begin{multline*}
    \funcstyle{even}(0), \funcstyle{even}(\funcstyle{s}(\funcstyle{s}(N))) \lrule \funcstyle{even}(N), \\
    \funcstyle{all\_even}([]), \funcstyle{all\_even}(A \cdot B) \lrule \funcstyle{even}(A), \funcstyle{all\_even}(B).
\end{multline*}

One could query $P$ either with the goal $\funcstyle{all\_even}(X)$ to find some $X$ such that all elements of $X$ are even or with the goal $\funcstyle{all\_even}([\funcstyle{s}(\funcstyle{s}(0))])$ to check whether all elements of the given list are even.

Reasoning about such queries can be implemented using the \emph{resolution rule}, an inference rule with appealing soundness and completeness properties \cite{Resolution}.
The resolution rule allows one to replace a rule's body atom by another rule's body, provided that rule's head unifies with the respective body atom.
Formally, for a substitution $\theta$:
\[
    \infer{\theta(h \lrule p_1, \dots, p_{i-1}, q_1, \dots, q_n, p_{i+1}, \dots, p_m)}{%
            h \lrule p_1, \dots, p_i, \dots, p_m
        &   h' \lrule q_1, \dots, q_n
        &   \theta(h') = \theta(p_i)
    }
\]

One can use the resolution rule to solve a goal $G$ by contradiction.
For example, considering $G = \funcstyle{all\_even}([\funcstyle{s}(\funcstyle{s}(0))])$ from the example above, one would negate the goal $G$ to obtain the query $Q = \bot \lrule G$, and attempt to derive a contradiction using the resolution rule within the Horn clause theory $P \cup \{ Q \}$.
A proof for this goal is depicted in \cref{fig:resolution-exm}.

Powerful programming languages have been built using the resolution rule or variants thereof, for example, Prolog \cite{LogicProgramming} and Datalog \cite{Datalog}.
Concrete implementations of programming languages typically do not solve queries by searching for proofs of goals such as in \cref{fig:resolution-exm} but use methods like SLD resolution \cite{LogicProgramming}.
This typically leads to programs having multiple semantics.
Logical semantics interpret logic programs in first-order logic whereas procedural semantics interpret them by reference to the answer search that solvers employ.

\begin{figure*}
    \centering
    \[
        \infer{\bot \lrule \top}{%
            \infer{\funcstyle{all\_even}([])}{}
            & \infer{\bot \lrule \funcstyle{all\_even}([])}{%
                \infer{\funcstyle{even}(0)}{}
                & \infer{\bot \lrule \funcstyle{even}(0), \funcstyle{all\_even}([])}{%
                    \infer{\funcstyle{even}(\funcstyle{s}(\funcstyle{s}(0))) \lrule \funcstyle{even}(0)}{}
                    & \infer{\bot \lrule \funcstyle{even}(\funcstyle{s}(\funcstyle{s}(0))), \funcstyle{all\_even}([])}{%
                        \infer{\funcstyle{all\_even}([\funcstyle{s}(\funcstyle{s}(0))]) \lrule \funcstyle{even}(\funcstyle{s}(\funcstyle{s}(0))), \funcstyle{all\_even}([])}{}
                        & \infer{\bot \lrule \funcstyle{all\_even}([\funcstyle{s}(\funcstyle{s}(0))])}{}
                    }
                }
            }
        }
    \]
    \caption{Example proof by contradiction using the resolution rule.
    We do not explicitly state the respective unifiers but apply them directly.
    We terminate proof search whenever we apply a rule that is included in the logic program $P$ modulo unification.}
    \label{fig:resolution-exm}
\end{figure*}

\subsubsection{Constraint Domains}

In this work, we will use a more general version of logic programs than described above, namely \acrfullpl{CLP}.
In \glspl{CLP}, one distinguishes between predicates that have an interpretation in the logic program, i.e., they are defined by the program's Horn clauses, and predicates that have a pre-defined interpretation in a \emph{constraint domain} $\C$.
The latter type of predicates are \emph{constraints}, and a \emph{solver} solves sets of constraints.
\Glspl{CLP} support incomplete solvers, i.e., for a set of constraints a solver may return $\top$ (true), $\bot$ (false), or undefined.

\Glspl{CLP} provide two advantages over ordinary logic programs.
First, for certain problem domains, e.g., the real numbers, a solver can find solutions much more efficiently than a logic programming engine that just uses the resolution rule.
For example, we will use constraints to model signature verification and hash digest comparisons.
Second, \glspl{CLP} allow logic programs to capture infinitely many solutions for a goal, described by a constraint set.

The constraint domain consists of
\begin{enumerate*}[label=(\alph*)]
    \item a constraint signature $\Sigma_\C$, which at least must include $(=)$ as a constraint,
    \item a class of constraints $\formulas_{\Sigma_\C}$, a set of formulas over $\Sigma_\C$,
    \item a domain of computation $D_\C$ interpreting the constraints,
    \item a constraint theory $T_\C$ specifying the semantics of the constraints, and
    \item a solver $\solve_\C : 2^{\formulas_{\Sigma_\C}} \pto \{ \top, \bot \}$.
\end{enumerate*}
We require that the domain of computation is a model of the constraint theory, and that a constraint domain's solver is \emph{sound} in that it returns $\bot$ only if the set of constraints indeed is contradictory, i.e., if $\solve_\C(c) = \bot$ then $T_\C \models \neg \exists \vec{v} \bigwedge c$ where $\vec{v}$ binds all free variables occurring in $c$.
Practically, a solver should also be sound in that it returns $\top$ only if the constraints are satisfiable, but this is irrelevant to the semantics of \glspl{CLP}, as will become clear later.

We call those atoms interpreted in the logic program simply atoms and call atoms that are part of the constraint domain \emph{constraints}.

In \glspl{CLP}, proofs are encoded as \emph{derivations}.
A derivation is a sequence of atoms and constraints $\langle G_1 \mid c_1 \rangle \Rightarrow \langle G_2 \mid c_2 \rangle \Rightarrow \dots$.
The first element of a derivation step, $G_i$, is the current list of target atoms, or the \emph{goal}, and the second element $c_i$ is a set of constraints, understood as the conjunction of its elements.
To update a derivation step $\langle G_i \mid c_i \rangle$ to $\langle G_{i+1} \mid c_{i+1} \rangle$, one can either:

\begin{itemize}
    \item Apply the resolution rule and select an atom $p_j$ in $G_i$ and a rule $h \lrule b_1, \dots, b_m$ in the logic program, and then define $c_{i+1} = c_i$ and:
    \[
        G_{i+1} = (p_1, \dots, p_{j-1}, p_j = h, b_1, \dots, b_m, p_{j+1}, \dots, p_n).
    \]

    Note that unification is not performed directly, but is represented by the constraint $p_j = h$.

    \item Select a constraint $p_j$ in $G_i$, for which $\solve_\C(c_i \cup \{ p_j \}) \neq \bot$, in which case the derivation's next element is:
    \[
        \langle p_1, \dots, p_{j-1}, p_{j+1}, \dots, p_n \mid c_i \cup \{ p_j \} \rangle.
    \]
    \item Select a constraint $p_j$ in $G_i$, but $\solve_\C(c_i \cup \{ p_j \}) = \bot.$ In this case, the derivation's next element is $\langle \emptyset \mid \bot \rangle.$
\end{itemize}

A derivation is \emph{from} a goal $G$ if its first element is $\langle G \mid \emptyset \rangle$.
A derivation is \emph{successful} if its last element is $\langle \emptyset \mid c \rangle$ and $c \neq \bot.$
We extend the definition of facts to \glspl{CLP} in that a fact is any rule of the form $h \lrule c_1, \dots, c_n$ such that $c_1, \dots, c_n$ are constraints.
A set of constraints $c$ is an \emph{answer} for a goal $G$ in a \gls{CLP} $P$ if there exists a successful derivation from $G$ that ends in $\langle \emptyset \mid c \rangle$.

\subsection{The DNS}
\label{sec:dns-bg}

\paragraph{Overview}
The \gls{DNS} is a distributed database that associates typed data with hierarchical names.
Names are sequences of one or more labels separated by dots and terminated with the empty root label, e.g., \name{example.com.} (note the trailing dot separating the label ``com'' from the empty root label).
The name space forms a tree, which has the empty label at its root.
Note that in practice, the root label is often omitted and names like \name{google.com} are understood as equal to \name{google.com.}.

Data is represented by \glspl{RR} of different types, and the name that is associated with some \gls{RR} is that \gls{RR}'s \emph{owner name}.
A name can have multiple \glspl{RR} of the same type, and \glspl{RR} with the same owner name and of the same type form a so-called \emph{RRset}.

The \gls{DNS} name space is divided into \emph{zones}, which are sub-trees of the \gls{DNS} name space.
\emph{Name servers} serve the contents of one or more zones.
Name servers use \emph{zone cuts} to delegate a sub-tree of the name space to a different name server.
For example, the root zone, the root of which is the name ``\name{.}'', may cut the name space at each \gls{TLD}, e.g., \name{com.}, \name{net.}, etc.
The name at the root of a zone is that zone's \emph{apex}.

Clients typically use \emph{recursive resolvers} to query the \gls{DNS}, which in turn query name servers known to them.
When a recursive resolver queries a name server, that name server may respond either
\begin{enumerate*}[label=(\alph*)]
  \item with records that answer the query directly,
  \item with structural records that inform further recursive resolution, or
  \item with an error.
\end{enumerate*}
In the first case, a recursive resolver will forward the data to the requesting client, and in the second and third cases, the recursive resolver may send follow-up queries to other name servers.

The \gls{DNS} is described in numerous \glspl{RFC} and has evolved considerably over the recent decades.
One of the most substantial extensions is \gls{DNSSEC}, which supports the authentication of zone contents by specifying how name servers can sign the contents of the zones they administer.
At zone cuts, name servers can indicate which key material may be used to authenticate the child zone.
This allows a resolver to validate a zone's authenticity as long as it can authenticate one of the root zone's verification keys and as long as every zone cut indicates a key for the child zone.

\paragraph{RR Types}
We distinguish between \emph{data \gls{RR} types} and \emph{structural \gls{RR} types}.
There are many data \gls{RR} types, e.g., \A and \AAAA records for IP addresses, \TXT for arbitrary text data, and \MX records for mail servers.
More interesting, though, are structural records.
The \NS record signals a zone cut, indicating which name server is responsible for the child zone, and the \CNAME and \DNAME records indicate aliases.
While \CNAME records indicate direct aliases, e.g., \name{a.com.} may alias to \name{b.com.}, a \DNAME record maps all names below a name to the corresponding name below the target.
For example, a \DNAME record for \name{a.com.} with the value \name{b.com.} would map a query for \name{x.y.z.a.com.} to \name{x.y.z.b.com.}.

\Gls{DNSSEC} and its further extensions added more structural \gls{RR} types.
\RRSIG records provide signatures over data, \DNSKEY records provide the signature verification keys, and \DS records contain digests of child-zone verification keys.
\Gls{DNSSEC} also supports the authenticated denial of existence of records using the \NSEC and \NSECIII records.

\paragraph{The Wildcard Label}
Another source of complexity for the \gls{DNS} is the \emph{wildcard label} ``$\wc$.''
Conceptually, the wildcard label is simple.
When a record owner name uses a wildcard, e.g., \name{\wc.example.com.}, responses are synthesized for all names that match this name, e.g., \name{sub.sub.example.com.}.
The wildcard label semantics, however, are complex and subtle and an entire \gls{RFC} was dedicated to clarifying them \cite{DNSWildcards}.
A \emph{wildcard owner name} is a name that has the wildcard as its leftmost label.
Name servers will synthesize responses for records with a wildcard owner name whenever
\begin{enumerate*}[label=(\alph*)]
  \item the query name \emph{matches} the wildcard record owner name, and
  \item no synthesized part of the query name \emph{exists}.
\end{enumerate*}

Matching is defined as expected.
For a wildcard owner name ``\name{*.parent}'', any subdomain of ``\name{parent}'' matches that wildcard owner name.
A name is said to exist if it has records associated with it, or it has a child that has records associated with it.
For example, when a zone contains records for \name{x.y.example.}, the names \name{x.y.example.}, \name{y.example.}, \name{example.}, and \name{.} all exist.
Considering \name{*.example.} and \name{x.y.example.}, no response would be synthesized for a name containing \name{y.example.}, as that name exists, even though no records are associated with \name{y.example.}.

\section{Formal Model}
\label{sec:model}
We now describe how we formally model the \gls{DNS} to reason about the resolvability of zones, in particular to determine \emph{whether} a recursive resolver could find an answer in a given set of zone files.
Thus, we abstract from malicious name server behavior and transport issues and assume that as long as a recursive resolver can reach a name server, the name server will share all data relevant to a query.

To model the \gls{DNS}, we compile a zone configuration, a set of initial name servers, and a (possibly empty) set of trusted \DS{} records that bootstrap \gls{DNSSEC} validation to a \gls{CLP}, as described in \cref{sec:bg-lp}.
At a high level, our compilation procedure creates a ``base case'' fact for every RRset in a given zone, modeling that if a recursive resolver queried the name servers serving this RRset for exactly the RRset's type and owner name, the name servers would respond with that RRset.
For the structural record types \NS{}, \CNAME{}, \DNAME{}, and \RRSIG{}, we additionally compile ``step case'' rules that model how these record types would lead to follow-up queries by a recursive resolver.

In the base case compilation, we treat the RRsets that are part of the zone configuration as unauthenticated.
Our compiler also adds base case rules for the trusted \DS{} records and treats them as authentic, thereby bootstrapping \gls{DNSSEC} validation.

We compile the structural record types \NS{}, \CNAME{}, \DNAME{}, and \RRSIG{} to step case rules as follows.
The rules for \NS{} records model the propagation of responses from a name server serving a delegated child zone to the name server serving the parent zone.
The rules for \CNAME{} and \DNAME{} records model the propagation of responses from queries for the target name to queries for the alias.
Finally, the rules for \RRSIG{} records can turn unauthenticated responses into authenticated ones.

We describe this compilation procedure in the following paragraphs.
We start by describing how we represent zone data itself.
The representation that we choose can be easily obtained from a zone's textual representation.
Next, we describe how we encode names and data as ground terms.
Finally, we describe how we compile RRsets to Horn clauses.

\subsection{The DNS}


We represent a set of zones as follows.
We consider a finite set of labels $\labels$ containing the empty label $\epsilon$ and the wildcard label ``$\wc$''.
A \emph{name} over $\labels$ is a non-empty sequence of labels, separated by dots ``$\mathdot$'' and terminated by $\epsilon$.
We use $\cdot$ to denote the concatenation of names, which adds a ``$\mathdot$'' between consecutive labels.
For example, $\name{example.com.} = \name{example} \cdot \name{com} \cdot \epsilon$.

\begin{definition} \label{def:names}
  The set of \emph{names} is the smallest set  $\names$ such that:
  \begin{itemize}
    \item $\epsilon \in \names$, and
    \item if $n \in \names$ and $l \in \labels \setminus \{ \epsilon\}$, then $l \cdot n \in \names$.
  \end{itemize}
\end{definition}

The \gls{DNS} associates names with typed data.
We use the set of symbols $\types$ for the relevant types:
\begin{multline*}
  \types = \{ \A, \AAAA, \NS, \CNAME, \DNAME, \RRSIG, \\
    \DNSKEY, \DS, \NSEC, \NSECIII, \dots \}.
\end{multline*}
Here, we only list types that inform recursive resolution.
There are more types, e.g., \MX{} for email communication, which we do not list here.
\felix{Could list all in Appendix.}

We represent the data itself as complex terms where we use the corresponding symbol in $\types$ as a function symbol and the function arguments represent the individual fields of the data.
For example, an \NS{} record delegates the resolution of its owner name to a server $s$ and has one field.
The data of this \NS{} record would be represented as $\NS(s)$.
An \RRSIG{} record stores the signature for a set of other resource records, and thus has many fields, including a tag for the signing key, a reference to a signing algorithm, the signature itself, and more.
We use $\data$ to refer to the set of all data.

We identify zones by their apex in $\names$ and represent them as a partial function $z : \names \pto \types \pto 2^\data$, which associates names and types with the respective RRsets.
Zones are served by different name servers.
A \emph{zone configuration} $c$ specifies which name servers serve which zones, and $c$ is a partial function $c : 2^\names \pto \names \pto \names \pto \types \pto 2^\data$ where $c(s, n) = z$ means that name server set $s$ serves zone $z$, which has the apex $n$.

\david{Cool. Note if reader not familiar with currying and types they may struggle.
You could create a typed z = $N \pto T \pto 2^\data$ that you use twice in last paragraph.}

\subsubsection{Zone Well-Formedness}
\label{sec:dns-wellformed}

We assume that zone files are well-formed in that they adhere to basic \gls{RFC} guidelines, which DecDNS checks before compiling them.
For example, \gls{RFC} 1034 allows at most one \CNAME{} record associated with an owner name \cite[Sec.~3.6.2]{DNS1}.
Critically, the well-formedness checks require no advanced reasoning and are always local to each zone.
See \cref{apx:wellformedness} for their full description.

\subsection{Modeling Names and Data}
\label{sec:wc-model}

\subsubsection{Representing Names}

We represent names as lists of labels, and labels in $\labels$ as constants.
Effectively, we assume that $\labels \subseteq \consts$.
We define the translation function $\namerep : \names \cup \names^\wc \to \ts_\Sigma$, which translates names into lists of labels, i.e., terms over $\Sigma$.
We omit the empty label as it is implicit.\footnote{%
  By \cref{def:names}, every name in $\names$ is terminated by the empty label.
}
The root name ``.'' is thus represented by the empty list.
\[
  \namerep(l_1 \cdot \tdots \cdot l_n \cdot \epsilon) =
  \begin{cases}
    [ l_n, \dots, l_2, \wc ] & l_1 \in \labels^\wc \\
    [ l_n, \dots, l_1 ] & \text{otherwise.}
  \end{cases}
\]

We homomorphically extend $\namerep$ to sets of names, representing the set as a list, ordered by the canonical ordering of names as specified in \gls{RFC} 4034 \cite{DNSSECSyntax}.

Representing names as lists allows unification to be used for synthesis, \DNAME{} rewrites, and checking whether one domain is a subdomain of another.
For example, checking whether the name \name{a.example.com.} is a subdomain of \name{example.com.} can be easily realized by checking whether the lists $[ \name{com}, \name{example} \mid \_ ]$ and $[ \name{com}, \name{example}, \name{a} ]$ unify.
Note that the latter list corresponds to $\namerep(\name{a.example.com.})$, but the former one does not correspond to $\namerep(\name{example.com.})$.

\paragraph{Semantics of Wildcard Labels}
Due to wildcard labels, \gls{DNS} zones can effectively be very large.
In a naïve model, one could represent the semantics of wildcard owner names by the set of all ground names for which wildcard synthesis would occur, though such sets would be unwieldily large.

To represent the semantics of wildcard owner names more efficiently, we use sets of \emph{wildcard constraints}, which represent labels for which wildcard synthesis does \emph{not} occur (see \Cref{sec:dns-bg}).
These sets replace wildcard labels and thus form their own kind of labels.
\begin{definition}
  The set of \emph{wildcard-constraint labels} $\labels^\wc$ is defined as:
  \[
    \labels^\wc = \{ S \cup \{ \wc \} \mid S \subseteq \labels \}.
  \]
\end{definition}

Observe that every wildcard-constraint label contains the wildcard label $\wc$ itself.
This is because the literal wildcard owner name, by definition, always exists.
Thus, no name server would perform synthesis for a name containing the wildcard owner name.

We extend the set of names $\names$ to $\names^\wc$, which will use labels from $\labels^\wc$ in their leftmost label.

\begin{definition}
  \label{def:names-wc}
  The set of \emph{wildcard-constraint names} is the smallest set $\names^\wc$ such that:
  \begin{itemize}
    \item Let $n = l_1 \cdot \tdots \cdot l_k \in \names$.
    If $l_1 \neq \wc$, then $n \in \names^\wc$.
    \item Let $\wc \cdot l_2 \cdot \tdots \cdot l_k \in \names$ and $l \in \labels^\wc$.
    $l \cdot l_2 \cdot \tdots \cdot l_k \in \names^\wc$.
  \end{itemize}
\end{definition}

For example, assuming that only the name \name{y.example.} exists below \name{example}, the name \name{$\wc$.example.}  would be represented as $\{ \name{y}, \wc \} \name{.example.}$.

We lift zone configurations $c$ from $2^\names \pto \names \pto \names \pto \types \pto 2^\data$ to $2^\names \pto \names \pto \names^\wc \pto \types \pto 2^\data$ by replacing every name in the third argument, representing an RRset owner name, with the corresponding name in $\names^\wc$, following \cref{def:names-wc}.
The concrete constraints for a wildcard owner name $\wc \cdot n$ can easily be computed within $c$.

Whenever the wildcard synthesis of responses must be considered in our formal model, we represent wildcard owner names, i.e., names where $l_1 \in \labels^\wc$, by lists in which we represent the wildcard label with a variable $v \in \vars$.
Whenever we use such lists in \gls{CLP} rules, we represent the wildcard constraints $l \in l_1$ by inequality constraints $v \neq l$ in the rule body.
For example, consider again the example wildcard-constraint name $\{ \name{y}, \wc \} \name{.example.}$.
In a \gls{CLP} rule, we would model when wildcard synthesis occurs by a rule of the following structure:

\[
    p([\name{example}, V \mid \_ ]) \lrule V \neq \name{y}, V \neq \wc.
\]

More formally, we define the function $\namerep_{v,w} : \names \cup \names^\wc \to \ts_\Sigma$ that replaces wildcard-constraint labels with a variable $v \in \vars$ and uses a variable $w \in \vars$ for the list tail:
\[
  \namerep_{v,w}(l_1 \cdot \tdots \cdot l_n \cdot \epsilon) = \begin{cases}
    [ l_n, \dots, l_2, v \mid w ] & l_1 \in \labels^\wc \\
    [ l_n, \dots, l_1 ] & \text{otherwise.}
  \end{cases}
\]

The function $\wcc_v : \names \cup \names^\wc \to 2^{\formulas}$ associates a name with a set of constraints for $v$ like those occurring in the rule body above.
The constraints represent all labels for which no synthesis will occur:
\[
  \wcc_v(l_1 \cdot n) = \begin{cases}
    \{ v \neq l \mid l \in l_1 \} & l_1 \in \labels^\wc \\
    \emptyset & \text{otherwise}.
  \end{cases}
\]

\subsubsection{Representing Data}

To represent data, we introduce the function $\datarep : 2^\data \to \ts_\Sigma$.
For an RRset $\{ f(x_1, \dots, x_n), \dots \} \in 2^\data$, $\datarep$ maps that set to a list, representing each of the RRset's elements in no particular order, using the respective function term $f/n \in \funcs$, and representing the concrete data item $x_i$ either as an atom or as $\namerep(x_i)$, if $x_i$ is a name.

\david{Usually if you map a set to a list you need some ordering for this function to be well defined. A standard way is via some lexicographer order. It seems you really don't care what ordering is used, but that doesn’t excuse not saying how you solve this.}

\subsection{A Formal Model of Recursive Resolution}

In this section, we describe how we translate a zone configuration into a logic program that allows one to reason about the resolvable queries for that configuration.

\subsubsection{Logic Program Domain}

Our logic program uses the following function symbols $\funcs$:
\begin{itemize}
  \item $\cons/2$ and $\emptylist/0$ for lists.
  \item $\bot/0$ and $\top/0$ for a Boolean.
  \item $0/0$ and $1/0$ for a bit.
  \item $\referral/0$ to model when a name server serves non-authoritative data for referrals, e.g., \NS{} or glue records.
  \item All symbols in $\types$, with the respective arity of the data, to represent data.
\end{itemize}

We will use $x \notin \{ a, b, c, \dots \}$ as syntactic sugar for $x \neq a, x \neq b, x \neq c, \dots$.
Furthermore, our logic program will use the following predicate symbols $\preds$:
\begin{itemize}
  \item $\member/2$ for list inclusion,
  \item $\min/3$ for the conjunction of two AD bits,
  \item $\resolves/8$ to model recursive resolution considering response data, and
  \item $\resolvable/3$ to model which queries are resolvable without considering the data.
\end{itemize}

$\resolves(s,t,z,n,d,r,v,a)$ models that querying name server set $s$ with query type $t$ for query name $n$ results in the data $d$.
Furthermore, $z$ indicates the zone through which the answer was inferred, $r \in \{ \bot, \top \}$ indicates whether an alias was followed, $v \in \{ \bot, \top \}$ indicates whether the resolver is performing \gls{DNSSEC} validation, and $a \in \{ 0, 1 \}$ indicates whether the query answer could be validated as authentic.
Note that $v = \bot$ implies that $a = 0$; one cannot validate an answer as authentic without performing validation.
When $v = \top$, $a$ can be either $0$ or $1$.
If $a = 1$, the resolver validated the answer as authentic.
If $a = 0$, the resolver accepted the answer as insecure because it validated a delegation as insecure.
Note that $\resolvable(t,n,v)$ effectively is a projection of $\resolves(s,t,z,n,d,r,v,a)$, as will become clear later.

\subsubsection{Constraint Domain}
\label{sec:constraints}

To model the \gls{DNS}, we also use the following constraints:
\begin{itemize}
  \item $(\neq)/2$ for inequality,
  \item $\namec/1$ for whether a term corresponds to a valid name, e.g., checking that its length remains within the bounds of the \gls{DNS},
  \item $\digestValid/3$ and $\signatureValid/4$ for digest and signature verification, as well as
  \item $\nsecUnsigned/2$, $\closestEncloser/5$, and $\nsecIIIUnsigned/3$ for verifying proofs of insecure delegations.
\end{itemize}

We model $(\neq)$ and $\namec$ as constraints to represent large zones efficiently.
For example, some \DNAME{} and wildcard records can lead to a vast number of queries being resolvable, which we represent symbolically using the above two predicates.

We encode \gls{DNSSEC}-related checks using constraints because
\begin{enumerate*}[label=(\alph*)]
    \item we solve them only when they are instantiated on ground terms, and expressing them as \gls{CLP} constraints provides an efficient way to achieve that, and
    \item implementing, e.g., signature verification is much more straightforward in an ordinary programming language than in Horn clauses.
\end{enumerate*}

Formally, our logic program will use the constraint signature $\Sigma_\C = (\vars, \consts, \funcs, P)$, where $P$ contains the above predicates.
Our constraint domain's domain of computation $D_\C$ uses the universe of all ground terms.
It interprets $(\neq)$ as the complement of the identity relation, and it interprets all other constraints directly as agreeing with the solver implementation, which we will describe in \cref{sec:implementation}.
To obtain the constraint theory $T_\C$, we follow the straightforward construction from \cite{CLP}, and obtain it as the theory of all first-order formulae true in $D_\C$.

\david{Usually we talk about formulae being satisfiable (or valid) rather than ``true''. True is informal and even ambiguous (satisfiable versus valid).}
\felix{Good point, but that's what's written in \cite{CLP}. Need to check in detail.}

\subsubsection{The Formal Model}

To reason about the resolvability of a zone configuration $c : 2^\names \pto \names \pto \names^\wc \pto \types \pto 2^\data$, we compile the configuration's RRsets to a \gls{CLP} $\Gamma(s_0,k_0,c)$, where $s_0 \subseteq \names$ is a set of initial name servers and $k_0 \subseteq \data$ is a set of trusted \DS{} records that bootstraps \gls{DNSSEC} validation.
We assume that $k_0$ is for the apex of $c$'s topmost zone $z_0 \in \names$.
We compile every RRset into a ``base case'' rule, which models that if a recursive resolver queries a name server $s$ for some name $n$ and query type $t$, it will respond with the data that it has stored for that name and type, and that data will be unauthenticated.
Moreover, we compile every RRset with a structural record type into ``step case'' rules, which model how a recursive resolver interprets the respective structural record, e.g., leading to recursive queries.

Let $s \subseteq \names$ be a set of name servers, $z \in \names$ a zone name, $n \in \names^\wc$ an owner name, and $t \in \types$ a record type, such that $(s, z, n, t, d) \in c$.
In the following, we will define a function $\gamma_{s_0} : 2^\names \to \names \to \names^\wc \to \types \to 2^\data \to 2^\formulas$ that compiles $(s, z, n, t, d)$ to a set of Horn clauses.
The Horn clauses $\Gamma(s_0,k_0,c)$ associated with a zone configuration $c$ then include a rule $\funcstyle{R_\resolvable}$ that projects $\resolves$ to $\resolvable$ atoms, rules $A(s_0,k_0,z_0)$ for the trusted \DS{} records, and the images of $\gamma_{s_0}$ over $c$:
\begin{align*}
  \Gamma(s_0,k_0,c) = \{ \funcstyle{R_\resolvable} \} \cup & \; A(s_0,k_0,z_0) \\
    \cup & \bigcup_{(s, z, n, t, d) \in c} \gamma_{s_0}(s, z, n, t, d).
\end{align*}

We project $\resolves$ atoms to $\resolvable$ atoms, provided that the response was obtained from the initial name servers, not referral data, and that the query type was for a non-structural record type, i.e., not in the set:
\begin{multline*}
    \types_S = \{ \SOA{}, \NS{}, \CNAME{}, \DNAME{}, \DS{}, \DNSKEY{}, \\
        \NSEC{}, \NSECIII{}, \RRSIG{}, \NSECIIIParam{} \}.
\end{multline*}

We opted to exclude reasoning about the resolvability of structural record types because their purpose is not to be resolvable by themselves but to enable the (authenticated) resolvability of other records.

\begin{lprule}
    \funcstyle{R_\resolvable} = \resolvable(Q,N,V) \\
    Q \notin \types_S \\
    \resolves(\namerep(s_0),Q,Z,N,\_,\_,V,\_), Z \neq \referral
\end{lprule}

In the following, we will define $\gamma_{s_0}$ by a case distinction.
In each case, we will constrain which formulas must be in $\gamma_{s_0}$'s image.
The image of $\gamma_{s_0}$ is then understood to be the smallest set satisfying all constraints.
In some cases, parts of a rule will be similar.
To highlight them, we will underline those parts that differ.

\paragraph{Base Case Rules}

For every RRset in $c$'s image, we add a rule to represent that when we query a set of name servers for a query name and query type, they will respond with the RRset matching the query exactly.
We choose $\mathit{zone} \in \{ \referral, \namerep(z) \}$ according to whether the owner name $n$ is indeed part of zone $z$ or a referral.
As the name servers may perform synthesis, we add one rule covering a query for the literal wildcard owner name, and one rule considering synthesis.
Note that if $n$ is not a wildcard owner name, both rules will be equal.
\begin{multline*}
  \resolves(\namerep(s), t, \mathit{zone}, \namerep(n), \datarep(d), \bot, \bot, 0). \\
    \in \gamma_{s_0}(s, z, n, t, d)
\end{multline*}
\begin{multline*}
    \begin{lprule}
      \resolves(\namerep(s), t, \mathit{zone}, \namerep_{L,N}(n), \datarep(d), \bot, \bot, 0) \\
      \wcc_L(n), \namec(\namerep_{L,N}(n))
    \end{lprule} \\
    \in \gamma_{s_0}(s, z, n, t, d)
\end{multline*}

We compile the trusted \DS{} records similarly to the base case rules above.
If $k_0 = \emptyset$, then $A(s_0, k_0, z_0) = \emptyset$.
Otherwise, we compile the following rule to bootstrap \gls{DNSSEC} validation:
\begin{multline*}
  A(s_0, k_0, z_0) = \{ \resolves(\namerep(s_0), \DS, \namerep(z_0), \\
    \namerep(z_0), \datarep(k_0), \bot, \top, 1) \}.
\end{multline*}

\paragraph{Step Case Rules}

The compilation of structural records into ``step case'' rules depends on the record type.

\subparagraph{NS} Let $t = \NS$ and $d = \{ \NS(\vec{s}_1), \dots, \NS(\vec{s}_k) \}$, let $\vec{s} = \{ \vec{s}_1, \dots, \vec{s}_k \}$ be the set of name servers responsible for the child, and let $\namerep(n) = [ l_n, \dots, l_1 ]$ be the delegated name.

We compile \NS{} records into multiple rules whenever $n \neq z$, i.e., the \NS{} records are not at the zone's apex, and these rules can be grouped into two categories.
First, we compile rules that model the propagation of data from a child to the parent.
Intuitively, this models that one can use the \NS{} records from the parent zone to obtain data from the child zone, provided one can also obtain one of the respective servers' IP addresses.
We compile the following rule for each child name server $\vec{s}_i$, address record type $t' \in \{ \A, \AAAA \}$, and address lookup $\choices{glue}$, with one restriction, which we explain below.
\begin{multline*}
    \choices{glue} \in \{ \resolves(\underline{\namerep(s)}, t', \underline{\referral}, \namerep(\vec{s}_i), \_, \bot, \bot, 0), \\
        \resolves(\underline{\namerep(s_0)}, t', \underline{\_}, \namerep(\vec{s}_i), \_, \bot, \bot, 0) \}
\end{multline*}
\begin{multline*}
\begin{lprule}
  \resolves(\underline{\namerep(s)}, T, \underline{\namerep(z)}, [ l_n, \dots, l_1 \mid N ], D, R, V, A) \\
  \namec([ l_n, \dots, l_1 \mid N ]) \\
  \choices{glue} \\
  \resolves(\underline{\namerep(\vec{s})}, T, \underline{\namerep(n)}, [ l_n, \dots, l_1 \mid N ], D, R, V, A)
\end{lprule} \\
  \in \gamma_{s_0}(s, z, n, \NS, d)
\end{multline*}

Note that the name server address query $\choices{glue}$ does not follow aliases, which is required by RFC 2181 \cite{DNSClarifications}.

To model how resolvers obtain the name server's IP address, we distinguish whether the name server $\vec{s_i}$ sits below the delegated name $n$.
If $\vec{s}_i$ is at or below $n$, the recursive resolver expects the parent zone's name servers $s$ to serve glue records.
Thus, we only emit the rule that queries the name server address via the initial name servers $s_0$ if $\vec{s}_i$ is not at or below $n$.

Second, we compile rules that model the propagation of unauthenticated answers through insecure delegations while performing validation.
A resolver can establish that a delegation is insecure by validating an \NSEC{} or \NSECIII{} proof.
\NSEC{} and \NSECIII{} records provide authenticated denial of existence, and they indicate which types exist for a given owner name.
When they indicate that no \DS{} records exist at a zone cut, a resolver cannot obtain authentic child zone verification keys through the configured trust anchors, and can thus accept insecure responses \cite{DNSSECSyntax,DNSSECSemantics,NSEC3}.

The difference between \NSEC{} and \NSECIII{} records is that \NSEC{} records allow one to enumerate all names that occur in a zone, which can be prevented when using \NSECIII{}.
This is achieved by hashing the original owner names, which, though, changes how resolvers can authenticate that there is no \DS{} record at a zone cut.

There are three ways in which a resolver can learn that a delegation is insecure, and thus we compile the rule below for each choice of $\choices{insecure} \in \{ \mathit{nsec}, \mathit{nsec3}_a, \mathit{nsec3}_b \}$.
The rule models that if a resolver can validate a delegation to be insecure (modeled by $\choices{insecure}$) they will propagate insecure data as validated to be insecure.
Note that the validation flag changes from $\bot$ in the body to $\top$ in the head.
\begin{multline*}
\begin{lprule}
  \resolves(\namerep(s), T, \namerep(z), [ l_n, \dots, l_1 \mid N ], D, \bot, \underline{\top}, 0) \\
  T \neq \DS{} \\
  \namec([ l_n, \dots, l_1 \mid N ]) \\
  \resolves(\namerep(s), T, \namerep(z), [ l_n, \dots, l_1 \mid N ], D, \bot, \underline{\bot}, 0) \\
  \choices{insecure}
\end{lprule} \\
  \in \gamma_{s_0}(s, z, n, \NS, d)
\end{multline*}

In the model for $\choices{insecure}$, the $\resolves$ predicates obtain the records used to prove that the delegation is insecure, and the constraints $\nsecUnsigned$, $\nsecIIIUnsigned$, and $\closestEncloser$ model the necessary checks on those records.

\begin{multline*}
    \mathit{nsec} = \\
    \resolves(\namerep(s), \NSEC{}, \namerep(z), \namerep(n), [D_\NSEC], \bot, \top, 1), \\
    \nsecUnsigned(\namerep(n), D_\NSEC)
\end{multline*}

\NSECIII{} records have particular owner names, which hide the records' original owner names.
Namely, their owner names are the name of the zone with a hash prepended, for example, \name{S83{\dots}5D8.example.com.}.
For a parent zone with apex $z = l_1 \cdot \dots \cdot l_n$, we represent such names with the two lists below, each containing a variable for the hashed label:

\[
    n_1 = [l_n, \dots, l_1, L_1] \quad
    n_2 = [l_n, \dots, l_1, L_2]
\]
\begin{multline*}
    \mathit{nsec3}_a = \\
    \resolves(\namerep(s), \NSECIII{}, \namerep(z), n_1, [D_\NSECIII{}], \bot, \top, 1), \\
    \nsecIIIUnsigned(\namerep(n), n_1, D_\NSECIII{})
\end{multline*}
\begin{multline*}
    \mathit{nsec3}_b = \\
    \resolves(\namerep(s), \NSECIII{}, \namerep(z), n_1, [D_\NSECIII{}], \bot, \top, 1), \\
    \resolves(\namerep(s), \NSECIII{},\namerep(z),  n_2, [D_\NSECIII{}'], \bot, \top, 1), \\
    \closestEncloser(\namerep(n), n_1, D_\NSECIII{}, n_2, D_\NSECIII{}')
\end{multline*}


As per \cite{DNSWildcards}, the semantics of \NS{} records with a wildcard owner name are undefined, and we thus do not consider the wildcard synthesis of responses for \NS{} records.

\subparagraph{CNAME} Let $t = \CNAME$ and $d = \{ \CNAME(\vec{n}) \}$.
For \CNAME{} records, we model that if we can resolve the target name $\vec{n}$ starting from the initial name servers, we can also resolve the alias at the name servers $s$ serving the \CNAME{} record.
Note that there must be only one \CNAME{} record per owner name, which is why we only consider $d$ to contain one element above.
We forbid the resolution of \CNAME{} queries using this rule ($T \neq \CNAME{}$) as those queries would be answered by the corresponding \CNAME{} record directly, and forbid the resolution of \gls{DNSSEC}-related records $\typesSec$, as their resolution does not follow aliases:
\[
    \typesSec = \{ \DS{}, \DNSKEY{}, \NSEC{}, \NSECIII{}, \RRSIG{} \}.
\]
\begin{multline*}
\begin{lprule}
  \resolves(\namerep(s), T, \namerep(z), \namerep(n), D, \top, \bot, 0) \\
    T \notin \{ \CNAME{} \} \cup \typesSec \\
    \namec(\namerep(n)) \\
    \resolves(\namerep(s_0), T, Z, \namerep(\vec{n}), D, \_, \bot, 0) \\
    Z \neq \referral
\end{lprule} \\
  \in \gamma_{s_0}(s, z, n, \CNAME, d)
\end{multline*}

Similarly to our model of \NS{} records, we model that \CNAME{} records propagate the authentication status of aliases.
A response for a query to an aliased name can be authenticated if both the query for the \CNAME{} record and for the alias target can be authenticated.
If either query has an insecure response, the combined response is insecure as well, which we model using the variable $A$ and the $\min$ predicate below.
\begin{multline*}
\begin{lprule}
  \resolves(\namerep(s), T, \namerep(z), \namerep(n), D, \top, \top, A) \\
    T \notin \{ \CNAME{} \} \cup \typesSec \\
    \namec(\namerep(n)) \\
    \resolves(\namerep(s_0), \CNAME{}, \_, \namerep(n), \datarep(d), \bot, \top, A_1) \\
    \resolves(\namerep(s_0), T, Z, \namerep(\vec{n}), D, \_, \top, A_2) \\
    \min(A_1, A_2, A), Z \neq \referral
\end{lprule} \\
  \in \gamma_{s_0}(s, z, n, \CNAME, d)
\end{multline*}

\CNAME{} records may have wildcard owner names.
We thus also add rules to $\gamma_{s_0}(s, z, n, \CNAME{}, d)$ that consider the synthesis of responses.
We compile these rules just as before but with the difference that we add the constraints $\wcc_L(n)$ to the rule body and replace $\namerep(n)$ with $\namerep_{L,N}(n)$ outside the query for the validated \CNAME{} record.

\subparagraph{DNAME} Let $t = \DNAME,$ $d = \{ \DNAME(\vec{n}) \}, \namerep(n) = [ l_n, \dots, l_1 ],$ and $\namerep(\vec{n}) = [ \vec{l}_m, \dots, \vec{l}_1 ]$.
We compile \DNAME{} records analogously to \CNAME{} records, with the only difference being that names are rewritten according to the \DNAME{} rules \cite{DNAMERecord} and that we allow for resolving \DNAME{} queries.
We model the rewriting of \DNAME{} records using the variables $L$ and $N$, which capture the non-empty list suffix $[L \mid N]$, corresponding to the rewritten name prefix.
We add the following two rules, the latter of which again models the propagation of the alias's authentication status.

\begin{multline*}
    \begin{lprule}
        \resolves(\namerep(s), T, \namerep(z), [ l_n, \dots, l_1, L \mid N ], D, \top, \bot, 0) \\
            T \notin \typesSec \\
            \namec([ l_n, \dots, l_1, L \mid N ]) \\
            \resolves(\namerep(s_0), T, Z, [ \vec{l}_m, \dots, \vec{l}_1, L \mid N ], D, \_, \bot, 0) \\
            Z \neq \referral
    \end{lprule} \\
    \in \gamma_{s_0}(s, z, n, \DNAME, d)
\end{multline*}
\begin{multline*}
    \begin{lprule}
        \resolves(\namerep(s), T, \namerep(z), [ l_n, \dots, l_1, L \mid N ], D, \top, \top, A) \\
            T \notin \typesSec \\
            \namec([ l_n, \dots, l_1, L \mid N ]) \\
            \resolves(\namerep(s_0), \DNAME{}, \_, \namerep(n), \datarep(d), \bot, \top, A_1) \\
            \resolves(\namerep(s_0), T, Z, [ \vec{l}_m, \dots, \vec{l}_1, L \mid N ], D, \_, \top, A_2) \\
            \min(A_1, A_2, A), Z \neq \referral
    \end{lprule} \\
    \in \gamma_{s_0}(s, z, n, \DNAME, d)
\end{multline*}

We reject \DNAME{} records with wildcard owner names, following the recommendation in \cite{DNSWildcards}, and therefore need not consider them in our compilation.

\subparagraph{RRSIG} Finally, we model the validation of responses via DNSSEC, using \RRSIG{} records.
Let $t = \RRSIG$ and $d = \{ \mathrm{sig}, \dots \}$ where $\mathrm{sig}$ is \RRSIG{} data covering type $t_\mathrm{COV}$ and signed by the operator of zone $n_\mathrm{SIG}$.
We distinguish between validation of \DNSKEY{} and other record types as well as between the validation of records with wildcard owner names and other owner names.
The difference is that \DNSKEY{} RRsets contain their own verification key, while one must first obtain an authentic verification key for other record types.
If $t_\mathrm{COV} \neq \DNSKEY$ and the record does not have a wildcard owner name, we add a rule for each data item $\mathrm{sig}$ within $d$ that
\begin{enumerate*}[label=(\alph*)]
  \item looks up the data to be validated,
  \item looks up an authenticated verification key RRset,
  \item non-deterministically chooses one of the keys in the RRset with the $\member$ predicate, and
  \item verifies the signature with that key with the $\signatureValid$ constraint.
\end{enumerate*}
\begin{multline*}
\begin{lprule}
  \resolves(\namerep(s), t_\mathrm{COV}, \namerep(z), \namerep(n), D, \bot, \underline{\top, 1}) \\
    \resolves(\namerep(s), t_\mathrm{COV}, \namerep(z), \namerep(n), D, \bot, \underline{\bot, 0}) \\
    \resolves(\namerep(s), \DNSKEY, \namerep(z), \namerep(n_\mathrm{SIG}), D_\mathrm{KEY}, \bot, \top, 1) \\
    \member(K, D_\mathrm{KEY}) \\
    \signatureValid(\namerep(n), D, \mathrm{sig}, K)
\end{lprule} \\
  \in \gamma_{s_0}(s, z, n, \RRSIG, d)
\end{multline*}

If the \RRSIG{} record has a wildcard owner name, we generate the same rule as above but replace all occurrences of $\namerep(n)$ outside the $\signatureValid$ constraint with $\namerep_{L,N}(n)$ and add the corresponding constraints $\wcc_L(n)$ to the rule body.
While the respective \glspl{RFC} contain explicit mechanisms for signature verification for synthesized responses, we perform the signature check directly on the record owner name instead of the synthesized name.

Note that the validation of responses synthesized from a wildcard owner name also requires validating that no non-synthesized answer to the query exists, which we omit in our model for performance reasons.
If a direct answer to the query existed, the name server would not synthesize the response.
Thus, if a resolver omits the check of non-existence, an attacker could use the signature over a record with a wildcard owner name to ``overwrite'' the name server's actual response.

As \DNSKEY{} RRsets contain their own verification key, we cannot authenticate that key by performing another \DNSKEY{} lookup.
\DNSKEY{} verification keys are authenticated by authentic \DS{} RRsets, which replace the \DNSKEY{} lookup from the rule above in our model for \DNSKEY{} RRset validation.
Thus, if $t_\mathrm{COV} = \DNSKEY$, then we add the rule below for each data item $\mathrm{sig}$ within $d$.
\begin{multline*}
\begin{lprule}
  \resolves(\namerep(s), \DNSKEY, \namerep(z), \namerep(n), D, \bot, \underline{\top, 1}) \\
    \resolves(\namerep(s), \DNSKEY, \namerep(z), \namerep(n), D, \bot, \underline{\bot, 0}) \\
    \resolves(\namerep(s_0), \DS, \_, \namerep(n), D_\mathrm{DS}, \bot, \top, 1) \\
    \member(K, D) \\
    \member(K_\mathrm{DIG}, D_\mathrm{DS}) \\
    \digestValid(\namerep(n), K, K_\mathrm{DIG}) \\
    \signatureValid(\namerep(n), D, \mathrm{sig}, K)
\end{lprule} \\
  \in \gamma_{s_0}(s, z, n, \RRSIG, d)
\end{multline*}

We do not consider the case that a \DNSKEY{} RRset is synthesized from an RRset with a wildcard owner name.
This is because \DNSKEY{} RRsets must only occur at the apex of a zone, which requires delegation by an \NS{} record with the same owner and thus wildcard owner name, which has undefined semantics \cite{DNSWildcards}.

\section{Change-Impact Analysis}
\label{sec:reasoning}
In this section, we present an algorithm to compute a change-impact analysis for two \glspl{CLP}.
This algorithm proceeds in two steps.
First, it computes all facts that are a \emph{consequence} of the \gls{CLP}, i.e., everything that is resolvable in a given zone configuration.
Second, it computes the difference between the two sets of resolvable queries.
Note that this difference cannot be computed by a simple set difference as the computed facts may contain residual constraints, for example, representing wildcard synthesis.
We will prove soundness and completeness of our algorithm in \Cref{sec:soundness}.

\subsection{Resolvable Queries}

For a \gls{CLP} $P$, a fact $a \lrule c$ is an \emph{immediate consequence} \cite{CLP} of a set of facts $F$ using a rule $R = h \lrule b_1, \dots, b_n \in P$ if $a = h$ and there is a derivation $\langle b_1, \dots, b_n \mid \emptyset \rangle \Rightarrow \dots \Rightarrow \langle \emptyset \mid c \rangle$ in $F$.
Intuitively, a rule has an immediate consequence with respect to a set of facts $F$ if all predicates in the rule body but not the constraints can be unified consistently with some fact in $F$.
$S_R(F)$ denotes the set of $F$'s procedural immediate consequences using $R$, and $S_P(F)$ denotes a program's procedural immediate consequences, defined as:
\[
    S_P(F) = \bigcup_{R \in P} S_R(F).
\]

For a set of initial name servers $s_0$ and trusted \DS{} records $k_0$, all resolvable queries $Q_\resolvable(s_0, k_0, c)$ of zone configuration $c$ can then be computed as the least fixpoint $\lfp$ of the $S_P$ operator, i.e., as the $\subseteq$-smallest set of facts $F$ for which $S_P(F) = F$.
\[
    \begin{aligned}
        Q_\resolvable(s_0,k_0,c) &= \{ a \lrule \Psi \in \lfp(S_{\Gamma(s_0,k_0,c)}) \\
        &\qquad \mid a~\text{is a}~\resolvable~\text{atom} \}.
    \end{aligned}
\]

\subsection{Change-Impact Analysis}

The change-impact analysis for a zone configuration update can be computed from $Q_\resolvable$.
Given a zone configuration $c$ and its updated version $c'$, the goal is to compute the symmetric difference of all ground instances entailed by $Q_\resolvable(s_0,k_0,c)$ and $Q_\resolvable(s_0',k_0',c')$, using the initial name servers and trust anchors for each version.
Note that we cannot directly compute the symmetric difference of $Q_\resolvable$'s images as they contain facts with potentially different constraints.

We define the operator $(\setminus) : \formulas \pto \formulas \pto 2^\formulas$.
$\phi_1 \setminus \phi_2$ is a set of formulas entailing all ground instances of $\phi_1$ without the ground instances of $\phi_2$, provided that $\phi_1$ and $\phi_2$ are both facts.

In general, $(\setminus)$ returns a set of formulas, not a set of facts, and these formulas may not even be \gls{CLP} rules anymore.
Consider, for example, $p(X) \setminus p(f(Y))$.
The entailed ground terms are all terms that do not use $f$ as a top-level symbol in $p$'s first argument, i.e., $p(X) \lrule \neg \exists Y. p(X) = p(f(Y)),$ which can be rewritten to $p(X) \lrule \forall Y. p(X) \neq p(f(Y)).$

We thus model subtraction using universally quantified inequality constraints.
As $\forall$ is part of the constraint domain's constraint language, and $\neq$ is a constraint, our implementation will treat conjuncts as constraints, and simplify them following some simple heuristics.

When computing $\phi_1 \setminus \phi_2$ there are two cases.
Either a ground term is in $\phi_1$ but not in $\phi_2$ because it does not unify with $\phi_2$'s head, or, if it does unify with $\phi_2$'s head, because it violates one of the constraints in $\phi_2$.
In general, the result is a disjunction.
We define $(\setminus)$ in \cref{alg:disunif}.

\begin{algorithm}
  \begin{algorithmic}[1]
    \STATE Let $\phi_1 = h_1 \lrule \Psi_1$, $\phi_2 = h_2 \lrule \Psi_2$, $\theta = \MGU(h_1, h_2)$.
    \IF{no such $\theta$ exists}
        \RETURN $\{ \phi_1 \} $ \label{ln:cia-id}
    \ELSE
        \STATE Let $\hat{u} = \{ h_1 \lrule \{ \forall \varsf(h_2). h_1 \neq h_2 \} \cup \Psi_1 \}$.%
        \label{ln:cia-no-unif}
        \STATE Let $u = \{ \theta(h_1) \lrule \theta(\Psi_1) \cup \{ \theta(\neg c) \} \mid c \in \Psi_2 \}$.%
        \label{ln:cia-unify}
        \RETURN $u \cup \hat{u}$
    \ENDIF
  \end{algorithmic}
  \caption{Algorithm to compute $\phi_1 \setminus \phi_2$.
  We assume that $\phi_1$ and $\phi_2$ share no variables.}
  \label{alg:disunif}
\end{algorithm}

First, to capture that the heads do not unify, \cref{alg:disunif} states that they must not unify as a universally quantified inequality constraint in \cref{ln:cia-no-unif}.
Second, to capture when the heads unify but one of the constraints is false, we unify the heads and negate each constraint of $\phi_2$ in \cref{ln:cia-unify}.

We extend $(\setminus)$ to finite sets of formulas $\Gamma_1, \Gamma_2 \subseteq \formulas$ as follows:
\begin{gather*}
    \Gamma_1 \setminus \psi := \bigcup_{\phi \in \Gamma_1} \phi \setminus \psi \\
    \Gamma_1 \setminus \Gamma_2 := ((\Gamma_1 \setminus \psi_1) \setminus \dots) \setminus \psi_n, \quad \Gamma_2 = \{\psi_1, \dots, \psi_n\}
\end{gather*}

Thus, finally, we define the symmetric difference of ground instances entailed by two sets of facts as follows:
\[
    \Gamma_1 \symdiff \Gamma_2 := (\Gamma_1 \setminus \Gamma_2) \disjcup (\Gamma_2 \setminus \Gamma_1)
\]

\david{You should prove that order of subtraction doesn't matter.}
\codex{State whether order independence is semantic or syntactic and explain why the intermediate facts still meet the pairwise subtraction theorem's hypotheses. Also specify how atom-head inequalities fit the constraint signature, which currently defines terms and atoms separately.}


\section{Soundness and Completeness}
\label{sec:soundness}
In this section, we provide soundness and completeness results for the algorithms that we presented in the previous section.
We begin by showing that computing the set of resolvable queries as the least fixpoint of the immediate consequence operator indeed computes all queries that are resolvable within the \gls{CLP}.
To establish this, we can straightforwardly reuse results established in prior work \cite{CLP}.
Afterwards, we provide soundness and completeness results for our change-impact analysis.

\subsection{Resolvable Queries}

In this section, we prove that $Q_\resolvable(s_0,k_0,c)$ actually computes all $\resolvable$ facts derivable from the \gls{CLP} $\Gamma(s_0,k_0,c)$.
We do this by relating the least fixpoint of the $S_P$ operator to a logical interpretation of a \gls{CLP}, following prior work \cite{CLP}.

For a \gls{CLP} $P$, an atom $a$ is a \emph{logical immediate consequence} of a set of atoms $A$ if there is a rule $r = h \lrule b_1, \dots, b_n \in P$ and valuation $\theta$ such that $a = \theta(h)$ and both
\begin{enumerate*}[label=(\alph*)]
    \item for all atoms $b_i$ occurring in $r$'s body it holds that $\theta(b_i) \in A$, and
    \item for the conjunction $c$ of all constraint formulas occurring in $r$'s body it holds that $D_\C \models \theta(c).$
\end{enumerate*}
The logical immediate consequence function $T^\C_P(I)$ maps $I$ to all its logical immediate consequences for $P$.

\felix{\cite{CLP} only defines $T^\C_P$ for constraints in case (b), i.e., atom constraints not formulas.
I extend this definition here so that we can also apply it to the change-impact analysis.
Doesn't make a difference for resolvable queries because they only contain constraints anyway.
Will note this in Appendix.}

One can relate a program's logical immediate consequences to a program's immediate consequences as defined by $S_P$.
To do so, though, we must first interpret the constraints that remain when applying $S_P$, which we do with the operator  $[\;]_\C$.
$[F]_\C$ maps a fact $F$ to its set of ground consequences in the constraint domain.
Formally:
\[
    [h \lrule c_1, \dots, c_n]_\C = \{ \sigma(h) \mid D_\C \models_\sigma c_1 \land \dots \land c_n \}.
\]

We extend $[\;]_\C$ to sets of facts as expected:
\[
    [S]_\C = \bigcup_{F \in S} [F]_\C.
\]

It has been shown in \cite{CLP} that the least fixpoint of $S_P$, when its constraints are interpreted, is equal to the least fixpoint of $T^\C_P$, which interprets constraints directly in the constraint domain.
The \emph{least fixpoint} of a function $f$, denoted $\lfp(f)$, is the smallest set $M$ such that $M = f(M)$.

\begin{thm}[Theorem~4.8 in \cite{CLP}]
    \label{thm:sound-complete-resolvable}
    Let $P$ be a \gls{CLP}. Then:
    \[
        [\lfp(S_P)]_\C = \lfp(T^\C_P).
    \]
\end{thm}

\Cref{thm:sound-complete-resolvable} applies directly to $Q_\resolvable$, which only restricts $\lfp(S_{\Gamma(s_0,k_0,c)})$ to facts with particular heads.
It shows that the ground instances of the facts in $Q_\resolvable(s_0,k_0,c)$ are exactly the $\resolvable$ atoms in the logical least fixpoint of $\Gamma(s_0,k_0,c)$.
Thus, our algorithm is sound ($\subseteq$) and complete ($\supseteq$).

\subsection{Change-Impact Analysis}
\label{sec:cia}

We show soundness and completeness of our change-impact analysis again by reference to $T_P^\C$.
Concretely, we show that the immediate consequences of $\phi_1 \setminus \phi_2$ correspond to the immediate consequences of $\phi_1$ without the immediate consequences of $\phi_2$.
In the following, we assume that $\phi_1$ and $\phi_2$ do not share any variables, which we call \emph{variable-disjoint}, and that all free variables occurring in the formula body occur in the formula head, which we call \emph{constraint-closed}.

\begin{thm}
    \label{thm:sound-complete-cia}
    Let $\phi_1, \phi_2$ be two variable-disjoint and constraint-closed rules, whose bodies only contain constraint formulas. Then:
    \[
        \lfp(T_{\phi_1 \setminus \phi_2}^\C) = \lfp(T_{\phi_1}^\C) \setminus \lfp(T_{\phi_2}^\C)
    \]
\end{thm}

We prove \cref{thm:sound-complete-cia} in \cref{apx:cia-proofs}.
The proof follows the construction of $\phi_1 \setminus \phi_2$ as per \cref{alg:disunif}.

We do not establish that the results of $Q_\resolvable$ will be both variable-disjoint and constraint-closed.
However, we enforce both in our implementation.
Variable-disjointness can be easily achieved by renaming facts before computing $(\setminus)$, and constraint-closedness is achieved by the design of rules in $\Gamma(s_0,k_0,c)$.
We check constraint-closedness nevertheless.

\section{Implementation}
\label{sec:implementation}
We provide a complete implementation of both our formal model and reasoning, as described in \cref{sec:model} and \cref{sec:reasoning}.
DecDNS is written in Haskell and provides multiple modes.
It can \begin{enumerate*}[label=(\alph*)]
    \item reason about the resolvability of zone files,
    \item compute a change-impact analysis for two sets of zone files,
    \item use our formal model to recursively resolve a \gls{DNS} query with real \gls{DNS} name servers, and
    \item compile a set of zone files to a \gls{CLP} that can be loaded into SWI-Prolog \cite{SWIProlog}.
\end{enumerate*}

Our implementation shows that our formal model and reasoning are practical.
Moreover, the mode for querying live \gls{DNS} servers to resolve \gls{DNS} queries shows that our formal model is so accurate and detailed that it can interoperate with real \gls{DNS} servers.

DecDNS has three parts:
\begin{enumerate*}[label=(\alph*)]
    \item a compiler from \gls{DNS} zone data to a \gls{CLP},
    \item a reasoning component for computing a least fixpoint, performing resolution, and computing a change-impact analysis, and
    \item a constraint solver supporting the constraints described in \cref{sec:constraints}.
\end{enumerate*}
The implementation of the compiler and the reasoning is straightforward.
The constraint solver requires more attention, though, and we thus explain it in more detail below.

\subsection{Constraint Solver Implementation}
\label{sec:solver-implementation}

The constraint solver provides support for the inequality constraint $(\neq)$ (also universally quantified), which we describe in detail in \cref{apx:ineqs}, the constraint $\namec$, and the \gls{DNSSEC}-related constraints.
\Gls{DNSSEC}-related constraints are only solved when all their arguments are ground.
To avoid repeated, expensive cryptographic checks such as signature verification, our constraint solver caches the results of previously solved ground \gls{DNSSEC} constraints.
Checking \gls{DNSSEC}-related constraints currently does not consider the records' validity periods.

The $\namec$ constraint checks the minimum length in bytes of a name potentially containing variables and evaluates to $\bot$ when a label exceeds the maximum length of 63 bytes or the name exceeds the maximum length of 255 bytes.
It evaluates to $\top$ only for ground inputs of valid length.
We assign a variable name $X$ a minimum length of $1$ as it may be the root name which is encoded as a zero byte.
We assign a name represented by the list $[l_1, \dots, l_n \mid \dots ]$ a minimum length of $1$ (for the root label) plus the sum of the lengths of its elements $l_1, \dots, l_n$, which represent labels.
We assign no length to a variable list tail as it may be empty.
Labels represented by atoms are of length $1$ (for the length byte) plus the atom's length.
Labels represented by variables have a minimum length of $2$ (the length byte plus at least one character).

\subsection{DNS Integration}
\label{sec:querying-mode}

\david{6.2 is veeery cool. But you should say why. It is because your model is of very high fidelity: accurate and detailed enough that it is [for some/
all/... what is scope?] queries interoperable with the entry DNS infrastructure. I.e., it can be used just like any other standards-conform
resolver for query resolution. My view: the kinds of things you describe in (A) are less interesting than why this integration itself is valuable.
Indeed as is, there is the risk that a reader thinks: well, we have lots of DNS resolvers, why do we need another. Or: I bet it is much slower
than the resolver I use...}

DecDNS provides a mode in which it resolves a query using real \gls{DNS} name servers, for which it only needs a list of root name server addresses, root apex \NS{} records, and a set of trusted \DS{} records.
This mode uses the resolution rule as described in \cref{sec:logic-resolution} and works as follows.

We compile an initial \gls{CLP} from the root hints, which identify a set $s_0$ of root name servers, and the supplied trust anchors $k_0$.
Then, we attempt to solve the \gls{DNS} query for query type $\varstyle{qt}$ and query name $\varstyle{qn}$ by solving the \gls{CLP} query $\resolves(\namerep(s_0), \varstyle{qt}, \_, \namerep(\varstyle{qn}), D, \_, \top, \_),$ i.e., we check whether we can obtain some data $D$ at the root name servers for the given query type and name while performing validation, but disregarding the particular zone, whether an alias was followed, and the final authentication status, thus allowing both authenticated answers and answers accepted as insecure.

During resolution in the \gls{CLP}, we record all failed goals, i.e., those for which there is no solution.
These failed goals encode potential follow-up queries.
Concretely, for every failed goal, we extract the goal's name server set, query name, and query type, which encodes a query to one of the name servers for that query name and type.
Initially, the goal $\resolves(\namerep(s_0), \varstyle{qt}, \_, \namerep(\varstyle{qn}), D, \_, \top, \_)$ will fail, which triggers the first query to a root name server.

Once we obtain all failed goals, we sort them according to a simple heuristic and send the respective queries to the respective name servers.
For the example above, this would initially be a query to one of the root name servers, which would likely return referral data.
We compile this new data to further \gls{CLP} rules, add those to our \gls{CLP}, and attempt again to solve the initial \gls{CLP} query.
We iterate until we find an answer to the query.

We used an existing library for \gls{DNS} query encoding and parsing, but all recursive resolution logic is handled by the \gls{CLP} and the query sorting heuristic.

\subsection{SWI-Prolog Integration}

SWI-Prolog is a popular Prolog interpreter \cite{SWIProlog}, which also supports \glspl{CLP}.
We provide an integration for SWI-Prolog so that it can reason about \glspl{CLP} compiled by DecDNS.
SWI-Prolog already supports inequality constraints natively.
For all other constraints, we provide a Haskell foreign-function interface, which allows Haskell code to be called from C code.
For full SWI-Prolog integration, we provide a small C module that provides support for $\namec$ and \gls{DNSSEC}-related constraints by linking their interpretation to our Haskell code.

\section{Evaluation}
\label{sec:evaluation}

In this section, we present our evaluation of DecDNS.
First, we evaluate whether DecDNS's reasoning is faithful to the \gls{DNS} standards, both when reasoning offline about zone files in \cref{sec:compilation} and when querying the live \gls{DNS} infrastructure in \cref{sec:online-eval}.
Second, we show which kinds of configuration errors DecDNS can detect (\cref{sec:eval-errors}).
We discuss limitations of DecDNS in this regard later in \cref{sec:conclusion}.
Finally, we measure DecDNS's performance in \cref{sec:performance}.

\subsection{Static Evaluation}
\label{sec:compilation}

We use the Ferret dataset~\cite{kakarla2022scale} to evaluate whether our model of recursive resolution is faithful to the \gls{DNS} standards.
The Ferret dataset was created to test name server \gls{RFC} compliance and contains zones and queries to those zones that exhibit corner cases of the \gls{DNS}.
It does not, however, contain signed zones.
For each zone in the dataset, we analyze which queries are resolvable according to DecDNS and compare the results with those of BIND~\cite{bind9} and Knot~\cite{knotresolver}.

Concretely, for a given zone and set of queries $Q$ in the Ferret dataset, we test the following combinations and queries:
\begin{itemize}
    \item We consider the zone once unsigned and once signed.
    The Ferret dataset also includes some invalid zones, which we only consider unsigned.
    \item We load the zone into BIND or Knot as the authoritative name server.
    \item We resolve each query in $Q$ with BIND or Knot as the recursive resolver.
    Resolver queries are directly forwarded to the name server serving the respective zone.
    \item We also generate random queries and resolve them with either BIND or Knot as the recursive resolver.
\end{itemize}

It can occur that a name server rejects loading a zone because it considers it malformed.
This can also occur for DecDNS whenever well-formedness checks fail.
When a zone is loaded by both the respective name server and DecDNS, we check whether $Q_\resolvable$ contains a matching query if and only if the resolver returns a positive response, and we additionally compare \gls{DNSSEC} validation results and response data.
This comparison subsumes different reasons why a resolver may return no data, including an error such as \SERVFAIL{} or \NXDOMAIN{} and a successful response with no data.
Overall, we evaluated 140,572 queries over 26,246 zones in four configurations (BIND/Knot name server $\times$ BIND/Knot resolver).

\paragraph{Results}
We find that DecDNS's output generally agrees with the real resolver software.
We only find six types of discrepancies between DecDNS and the name servers or resolvers, which we accept as inconsistencies.
During the evaluation, we found a number of errors in DecDNS, e.g., inconsistent serialization of record data for signature verification, incorrectly treating referral data as an answer, and incorrect handling of \NS{} and \DNAME{} records with the same owner name.
We fixed these errors and iterated the evaluation until we decided that the remaining inconsistencies were acceptable.

\Cref{tab:mismatch} provides the details of the accepted inconsistencies.
The largest source of inconsistencies is that the BIND name server rejects zones with missing name server addresses, but DecDNS does not.
DecDNS assumes that the addresses of the name servers serving the common root of the input zones are available to the resolver.
Only if a parent zone and delegation were considered would the lack of name server addresses become visible in DecDNS's output.

DecDNS rejects zones that contain non-referral data at or below a delegation, which Knot and BIND permit (see \cref{apx:wellformedness}).
\Gls{RFC} 2181~\cite[Section~6.1]{DNSClarifications} only requires name servers to ignore such data at or below a zone cut.
We choose to be stricter than the \glspl{RFC} as we still argue that such data indicates a configuration error.
DecDNS also does not synthesize \CNAME{} records for queries that match \DNAME{} records, but instead encodes their semantics directly in the model of recursive resolution.

In some cases, DecDNS reported queries as being resolvable but the resolver did not provide a positive response.
Mostly, this was due to rewrite limits of the respective resolver.
For nine queries, though, Knot returned an error after it entered a loop in an attempt to resolve \DS{} follow-up queries.

We also identified what we believe to be a Knot name server bug.
For a few queries, Knot omitted the relevant \NSEC{} proofs for a signed response synthesized from a wildcard owner name, leading to a failure in resolution by BIND and Knot while DecDNS reported the queries as resolvable.

Finally, one zone configuration from the evaluation led to a timeout.
This zone allows for the resolution of approximately $2^{40}$ queries due to \DNAME{} rewrites, which cannot be represented efficiently with constraints.

\begin{table}[!bpt]
    \centering
    \begin{tabular}{p{0.6\linewidth} r r}
        \hline
        Mismatch attribution & Queries & Zones \\
        \hline
        BIND name server rejects \NS address configurations & 2,700 & 38 \\
        DecDNS rejects data at or below delegation & -- & 1,556 \\
        DecDNS omits synthesized \CNAME records & 776 & 45 \\
        Resolver alias rewrite limits & 259 & 7 \\
        Knot resolver loops on \DS lookups & 9 & 9\\
        Knot name server omits wildcard \NSEC proofs & 3 & 3 \\
        Timeout & -- & 1 \\
        \hline
    \end{tabular}
    \caption{Categories of observed differences.
    For query counts, categories may overlap.}
    \label{tab:mismatch}
\end{table}

\subsection{Online Evaluation}
\label{sec:online-eval}

We also evaluated DecDNS against real \gls{DNS} infrastructure using a ``100,000 queries'' file, which the BIND \cite{bind9} implementation provides and uses for A/B testing new releases.
From this file, we randomly sampled 2,775 queries for record types supported by DecDNS.
We resolved each query with both the \texttt{dig} utility and DecDNS in querying mode (see \cref{sec:querying-mode}), configured with a 30~s timeout and a limit of 100 queries to \gls{DNS} name servers.

We show our results in \cref{tab:online-eval}.
DecDNS timed out for approximately 67\% of the queries.
For 21\% of the queries, DecDNS matched the result from \texttt{dig}.
The results differed for around 12\% of the queries.

We attribute 20 of these mismatches to a name server returning different data to \texttt{dig} and DecDNS.
This is common practice, e.g., for services using content-delivery networks.
Both \texttt{dig} and DecDNS log all records used for resolution.
After studying these logs, we concluded that both interpreted the data correctly but arrived at different results.

In all other cases, we attribute the mismatch to DecDNS raising a benign error when \texttt{dig} succeeds in resolving the query.
In some cases, DecDNS received records that violated well-formedness invariants.
For example, it crashed upon receiving \texttt{. NS localhost.}.
In other cases, there was a network error in our testing environment affecting only DecDNS, or DecDNS exhausted its query limit.

We note, though, these results could be subject to a survivorship bias, as more complex queries are more likely to lead to a timeout but also more likely to exhibit diverging behavior between \texttt{dig} and DecDNS.
Nevertheless, this evaluation shows that our approach provides some interoperability with real \gls{DNS} infrastructure.

\begin{table}
    \centering
    \begin{tabular}{llr}
    \hline
    Category & Reason & Count \\
    \hline
    {\bfseries Success} & -- & 579  \\
    {\bfseries Timeout} & -- & 1854 \\
    \multirow{4}{*}{\bfseries Mismatch}
      & Different response & 20 \\
      & Well-formedness errors & 8 \\
      & Network error & 290 \\
      & Query limit exhausted & 24 \\
    \hline
    \end{tabular}
    \caption{Results of the online evaluation.}
    \label{tab:online-eval}
\end{table}

\subsection{Detecting Configuration Errors}
\label{sec:eval-errors}

We evaluated whether DecDNS can detect configuration errors by introducing them into a number of example zones.
Because DecDNS performs a change-impact analysis, it can detect configuration errors whenever they manifest as a change in resolvability.
This, though, means that detecting configuration errors may require interpreting the results of the change-impact analysis.

\Cref{tab:configuration-errors} summarizes example configuration errors that we can detect.
We can most easily detect errors introduced by changes that should preserve the resolvability of existing queries.
This includes, for example, introducing a new zone cut or a \gls{DNSSEC} key rollover, the kind of change that led to the DENIC outage \cite{DenicOutage}.
For many other configuration errors, detection depends on whether the change introducing them deletes records.
For example, when decommissioning name servers and moving the zone to new ones, the old name server addresses would be expected to become unresolvable, and thus there would be an expected loss of resolvability.

\begin{table*}[t]
  \centering
  \small
  \renewcommand{\arraystretch}{1.15}
  \begin{tabularx}{\textwidth}{lXc}
    \toprule
    Configuration error
    & Error description
    & Interpretation? \\
    \midrule
    Cyclic zone dependencies
    & Name server address resolution forms a cycle.
    & $\bullet$ \\

    \addlinespace
    Delegation inconsistency
    & The parent's delegation and the child's authoritative name server information disagree.
    & \cmark \\

    \addlinespace
    DENIC outage
    & During a rollover, multiple signing keys are used but only a subset is
      published as DNSKEY records, as in DENIC's May 2026 outage \cite{DenicOutage}.
    & \xmark \\

    \addlinespace
    Incomplete zone cut
    & A new zone cut is introduced but some records are omitted.
    & \xmark \\

    \addlinespace
    Lame delegation
    & A parent delegates a child zone to a name server that does not serve the zone.
    & $\bullet$ \\

    \addlinespace
    Missing glue
    & A parent delegates a zone to a name server but does not include required glue records.
    & $\bullet$ \\

    \addlinespace
    Rewrite blackholing
    & Alias rewriting terminates with a non-existent name.
    & $\bullet$ \\

    \addlinespace
    Rewrite loop
    & Alias rewriting forms a cycle.
    & $\bullet$ \\

    \addlinespace
    Wildcard shadowing
    & A newly added name blocks wildcard synthesis.
    & \xmark \\
    \bottomrule
  \end{tabularx}
  \caption{
    Configuration errors that a resolvability regression analysis can detect.
    ``Interpretation?'' marks whether detecting the configuration error requires interpretation of the output.
    This is not the case when a change introducing that error is generally not expected to change the resolvability of data queries.
    ``$\bullet$'' indicates that interpretation may only sometimes be required.
  }
  \label{tab:configuration-errors}
\end{table*}

\subsection{Performance}
\label{sec:performance}

\subsubsection{Static Evaluation}

For the static evaluation as described in \cref{sec:compilation}, we measured DecDNS's runtime per zone, including process startup and output collection, but excluding zone signing.
We exclude configurations that fail well-formedness checks or time out.
The average runtime is 38.0~ms for unsigned and 92.2~ms for signed zones, with medians of 31.0~ms and 76.0~ms, respectively.
The 95th-percentile runtimes for original and signed configurations are 64.8~ms and 107.8~ms, respectively.
To investigate outliers, we inspected the ten slowest signed and unsigned configurations, all of which contained \DNAME{} loops, producing hundreds of derived facts from only four or five records.
Note that the Ferret dataset is relatively homogeneous in structure, and that no zone configuration contains more than five records.

\begin{figure}[!bpt]
    \centering
    \includegraphics[width=1\linewidth]{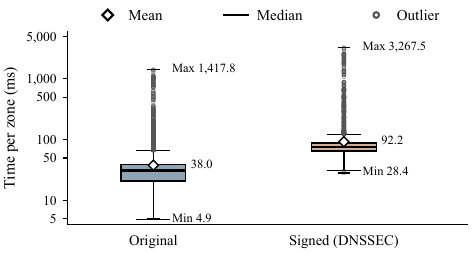}
    \caption{Runtime distribution per valid zone on a logarithmic scale. Boxes span the 25th--75th percentiles, and whiskers extend to minima and maxima excluding outliers.
    Outliers are shown as dots.}
    \label{fig:ferret_performance}
\end{figure}

\subsubsection{Online Evaluation}

\begin{figure}[!t]
    \centering
    \includegraphics[width=1\linewidth]{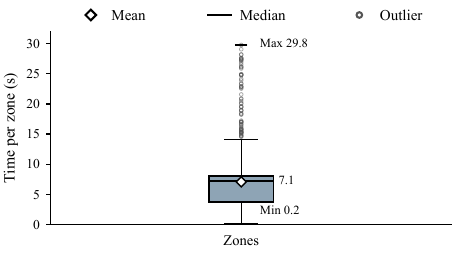}
    \caption{Runtime distribution for successful online queries, in seconds.
    Boxes span the 25th--75th percentiles, and whiskers extend to minima and maxima excluding outliers.
    Outliers are shown as dots.}
    \label{fig:online-eval-performance}
\end{figure}

We also timed resolution by DecDNS during the online evaluation described in \cref{sec:online-eval}.
\Cref{fig:online-eval-performance} shows the distribution of runtimes for queries that did not time out and did not raise an error.
The mean and median runtimes were both 7.1~s.

\section{Related Work}
\label{sec:rw}
Many works tackled the issue of establishing properties of \gls{DNS} zone configurations \cite{DelphiPaper,DNSMaude,GRoot,VeriDNS,DNSOctopus}.
All these works, though, lack support for \gls{DNSSEC} and suggest verifying the absence of particular types of configuration errors.
Many of these works provide support for verifying other properties, so it would theoretically be possible to verify different, more constructive properties, but no such properties are suggested.
In contrast, DecDNS performs a change-impact analysis, which allows for detecting configuration errors in a much more general way, and provides support for \gls{DNSSEC}.

There is also prior work that suggested the use of logic programming for reasoning about the \gls{DNS} \cite{SEMDNS}.
Their model, however, lacks many features that we cover.
For example, they do not cover \gls{DNSSEC}, the following of referrals, obtaining name server addresses, and \CNAME{} records with a wildcard owner name.
The authors also leave it open what their model could be used for and handle wildcard synthesis and \DNAME{} rewrites explicitly rather than symbolically as we do.

There are other works that studied \gls{DNSSEC} formally \cite{DNSSECVerif,DNSSECSecurityEval}, but these works studied the security of the \gls{DNSSEC} standards rather than the correctness of zones.
For example, both works analyzed whether an attacker can trick a resolver into accepting attacker-supplied data as valid.
Thus, these models are to some extent incomparable to ours.
Both works consider only a fixed example zone for their security analyses and do not consider aliases, and both analyses include a model of recursive resolvers.
The latter difference most clearly contrasts these works to ours.
We do not model recursive resolvers since our model \emph{is} the resolver.

Finally, a related but orthogonal work is ``Topaz'' \cite{Topaz}.
Topaz is a formally verified name server architecture that correctly serves data associated with domain names, and was designed for use within content-delivery networks.
Such networks often serve different IP addresses to clients, e.g., optimizing the server location with respect to the client location.
In Topaz, such optimizations can be expressed in a verified policy language which administrators can use to configure which \gls{DNS} data should be served to which clients.

\section{Discussion \& Conclusions}
\label{sec:conclusion}
In this paper, we presented a new formal model for reasoning about recursive resolution in the \gls{DNS}, as well as DecDNS, an implementation of that model.
DecDNS was designed to perform a change-impact analysis of zone configuration updates, which can help detect zone configuration errors.
Moreover, our formal model is so accurate and detailed that DecDNS can resolve \gls{DNS} queries with the real \gls{DNS} infrastructure.

\paragraph{Limitations}
Nevertheless, our work is not without limitations.
First, DecDNS's change-impact analysis does not detect all potential configuration errors.
\begin{enumerate*}[label=(\alph*)]
  \item DecDNS disregards the correctness of the data associated with names and thus cannot detect whether, e.g., the IP address associated with a name server is incorrect.
  \item DecDNS cannot detect when name servers serve inconsistent data.
  We assume that all name servers configured at a zone's apex indeed serve the respective zone.
  \item DecDNS does not detect when the name servers serving a configuration's topmost zone have no addresses configured.
  We assume resolvers are configured with those addresses.
  \item DecDNS reasons about whether a resolution path exists, which means that we may not detect that a particular resolution path, potentially taken by many resolvers, does not return a response.
\end{enumerate*}

Second, our model and algorithm have some practical limitations.
\begin{enumerate*}[label=(\alph*)]
  \item It could occur that our change-impact analysis returns formulas containing contradictory residual constraints that our constraint solver fails to recognize as such.
  We, however, never encountered such a case.
  \item DecDNS's performance can be further improved; for example, it timed out when we attempted to reason about the root zone.
  \item Our model currently omits some aspects of \NSEC{} and \NSECIII{} records, particularly when validating responses synthesized from signed wildcard records.
\end{enumerate*}

Third and finally, our evaluation does not cover \gls{DNSSEC} and multi-zone corner cases well.
The Ferret dataset (see \cref{sec:compilation}) includes only corner cases for individual, unsigned zones, and there is no characterization as to which configurations are covered by our online evaluation (\cref{sec:online-eval}).

\paragraph{Future Work}
We plan to address many of the above limitations in future work.
Concretely, we plan to improve performance, broaden the evaluation, and fully model \NSEC{} and \NSECIII{}.
Moreover, we believe that the change-impact analysis computed by DecDNS can also be generalized to not only consider old and new zone configurations but also different compilations of zone configurations.
For example, the same zone could be compiled for different versions of the \gls{DNS} standards or the compilation could model that all resolution paths must succeed instead of modeling the existence of a resolution path.

DecDNS's change-impact analysis currently requires complete knowledge of a zone and all its subzones.
One could investigate treating unresolved $\resolves$ goals for unknown zones as residual constraints, which could allow reasoning with partial knowledge.

Finally, it would be worthwhile to analyze DecDNS's security, particularly in querying mode.
Such an analysis would strengthen DecDNS's value as a potential reference implementation.

\section*{Acknowledgments}
We thank Steve Crocker, Petr \v{S}pa\v{c}ek, and Peter Thomassen for their interest in this project and their valuable feedback.


\bibliographystyle{plainurl}
\bibliography{references}

\appendix

\section{Proofs for Change-Impact Analysis}
\label[appendix]{apx:cia-proofs}
We begin our proof of \cref{thm:sound-complete-cia} by proving a lemma that allows us to rewrite our proof goal in simpler terms.

\begin{lem}
    \label{lem:fact-consequences}
    Let $F$ be a set of rules only containing constraint formulas in their body.
    Then $\lfp(T_F^\C) = T_F^\C(\emptyset)$.
\end{lem}

\begin{proof}
    Let $F$ be a set of rules only containing constraint formulas in their body.
    Observe that for such rules the definition of $T_P^\C$ is independent of its argument, that is:
    \[
        \forall x, y. T_F^\C(x) = T_F^\C(y)
    \]
    Thus, we obtain $\lfp(T_F^\C) = T_F^\C(\emptyset)$.
\end{proof}

We proceed to introduce two lemmas that relate $\lfp(T_{\phi_1 \setminus \phi_2}^\C)$ to $\lfp(T_{\phi_1}^\C)$ and $\lfp(T_{\phi_2}^\C)$, which will help prove that:
\[
    \lfp(T_{\phi_1 \setminus \phi_2}^\C) \subseteq \lfp(T_{\phi_1}^\C) \setminus \lfp(T_{\phi_2}^\C).
\]

\begin{lem}
    \label{lem:disunif-in-left}
    Let $\phi_1, \phi_2$ be two variable-disjoint facts.
    Then:
    \[
        \lfp(T_{\phi_1 \setminus \phi_2}^\C) \subseteq \lfp(T_{\phi_1}^\C).
    \]
\end{lem}

\begin{proof}
    Observe that in every case of \cref{alg:disunif}, we only strengthen $\phi_1$, i.e., either add more conjuncts to its body or additionally apply a substitution to its head.
    This directly proves \cref{lem:disunif-in-left}.
\end{proof}

\begin{lem}
    \label{lem:disunif-dif-right}
    Let $\phi_1, \phi_2$ be two variable-disjoint and constraint-closed facts.
    Then:
    \[
        \lfp(T_{\phi_1 \setminus \phi_2}^\C) \cap \lfp(T_{\phi_2}^\C) = \emptyset.
    \]
\end{lem}

We require that $\phi_1$ and $\phi_2$ are constraint-closed because \cref{alg:disunif} might otherwise return incorrect results for cases corresponding to \cref{ln:cia-unify}.
To realize that, consider the two facts $\phi_1 = p(X)$ and $\phi_2 = p(f(Y)) \lrule c(Z)$ for some constraint $c/1$.
Observe that $\phi_2$ is not constraint-closed.
For those formulas, we obtain:
\begin{gather*}
    \phi_1 \setminus \phi_2 = \{ p(X) \lrule \forall Y. p(X) \neq p(f(Y)), \; \psi \}, \\
    \psi = p(f(Y)) \lrule \neg c(Z).
\end{gather*}

As the heads of $\phi_2$ and $\psi$, which is obtained from \cref{alg:disunif} \cref{ln:cia-unify}, are independent of the constraint $c(Z)$ or $\neg c(Z)$ respectively, we can derive $p(f(Y))$ for any ground instance of $Y$ from both $\phi_1 \setminus \phi_2$ and $\phi_2$ provided that $c$ is neither tautological nor contradictory.
Clearly, this is not intended for $\phi_1 \setminus \phi_2$.
We thus limit the application of \cref{alg:disunif} to constraint-closed facts, which are the norm, and for which this problem does not arise, as we show below.

\begin{proof}
    Let $\phi_1, \phi_2$ be two variable-disjoint and constraint-closed facts such that
    \[
        \phi_1 = h_1 \lrule \Psi_1 \quad \phi_2 = h_2 \lrule \Psi_2
    \]

    Let $a \in \lfp(T_{\phi_2}^\C)$ and let $\sigma$ be a valuation such that $a = \sigma(h_2)$ and $D_\C \models \sigma(\Psi_2)$.
    Presume, for a contradiction, that $a \in \lfp(T_{\phi_1 \setminus \phi_2}^\C).$
    Since $a \in \lfp(T_{\phi_1 \setminus \phi_2}^\C)$, we have that $a \in T_{\phi_1 \setminus \phi_2}^\C(\emptyset)$ by \cref{lem:fact-consequences}, and thus, there must be a rule $\psi = h \lrule \Psi$ in $\phi_1 \setminus \phi_2$ such that $a \in T_\psi^\C(\emptyset)$, i.e., there must be a valuation $\sigma'$ such that $a = \sigma'(h)$ and $D_\C \models \sigma'(\Psi)$.
    We proceed with a case distinction.
    \begin{itemize}
        \item Let $\psi$ have the following shape, i.e., come from \cref{alg:disunif} \cref{ln:cia-no-unif}:
        \[
            h_1 \lrule \{ \forall \varsf(h_2). h_1 \neq h_2 \} \cup \Psi_1.
        \]

        Observe that:
        \[
            \sigma'(\forall \varsf(h_2). h_1 \neq h_2) = \forall \varsf(h_2). a \neq h_2.
        \]

        However, we have that $D_\C \not\models \forall \varsf(h_2). a \neq h_2$ as $a = \sigma(h_2)$.
        Thus, $\psi$ cannot be of the above shape.

        \item Let $\psi$ have the following shape for $\theta = \MGU(h_1, h_2)$ and $c \in \Psi_2$, i.e., come from \cref{alg:disunif} \cref{ln:cia-unify}:
        \[
            \theta(h_1) \lrule \theta(\Psi_1) \cup \{ \theta(\neg c) \}
        \]

        Observe that by assumption we have:
        \[
            D_\C \models \sigma'(\Psi) = \sigma'(\theta(\Psi_1) \cup \{ \theta(\neg c) \}).
        \]

        Thus and in particular, we have that $D_\C \models \sigma'(\theta(\neg c))$ or equivalently $D_\C \not\models \sigma'(\theta(c)).$

        Observe that since $\theta(h_1) = \theta(h_2),$ we have that $\varsf(\theta(h_1)) = \varsf(\theta(h_2))$.
        Consequently, we have that $\psi$ is constraint-closed, as $\phi_2$ is constraint-closed and thus $\varsf(\theta(c)) \subseteq \varsf(\theta(h_2))$.

        As $a = \sigma(h_2)$ and $a = \sigma'(\theta(h_1))$, and because both $\phi_2$ and $\psi$ are constraint-closed, we can deduce that $\sigma(c) = \sigma'(\theta(c))$.
        This means, though, that all of the following statements are true, which is a contradiction:
        \[
            D_\C \models \sigma(c), \quad
            D_\C \not\models \sigma'(\theta(c)), \quad
            \sigma(c) = \sigma'(\theta(c)).
        \]
    \end{itemize}
    In consequence of the above case distinction, we can deduce that $a \notin T_{\phi_1 \setminus \phi_2}^\C(\emptyset),$ which contradicts our initial assumption.
\end{proof}

Before we proceed to our proof of \cref{thm:sound-complete-cia}, we prove a helpful property about most general unifiers and their relationship to valuations.

\begin{lem}
    \label{lem:extend-mgu}
    Let $a_1$ and $a_2$ be two variable-disjoint atoms and $\theta = \MGU(a_1, a_2)$.
    If there exist valuations $\sigma_1, \sigma_2$ such that $\sigma_1(a_1) = \sigma_2(a_2),$ then there also exists a valuation $\sigma$ such that $\sigma_1(a_1) = \theta\sigma(a_1) = \theta\sigma(a_2) = \sigma_2(a_2).$
\end{lem}

\begin{proof}
    Let $a_1, a_2, \sigma_1, \sigma_2$ and $\theta$ as per \cref{lem:extend-mgu}.
    Let $\theta'$ be a valuation such that:
    \[
        \theta'|_{\varsf(a_1) \cup \varsf(a_2)} = \sigma_1|_{\varsf(a_1)} \cup \sigma_2|_{\varsf(a_2)}.
    \]
    Observe that $\theta'$ is a unifier for $a_1$ and $a_2$.
    Thus, there must exist a valuation $\sigma$ such that $\theta' = \theta\sigma$, as $\theta$ is a most general unifier.
\end{proof}

\begin{proof}[Proof of \cref{thm:sound-complete-cia}.]
    Let $\phi_1, \phi_2$ be two variable-disjoint and constraint-closed facts such that:
    \[
        \phi_1 = h_1 \lrule \Psi_1, \quad \phi_2 = h_2 \lrule \Psi_2.
    \]

    Presume that the heads of $\phi_1$ and $\phi_2$ do not unify.
    Then:
    \[
        \lfp(T_{\phi_1 \setminus \phi_2}^\C) = \lfp(T_{\phi_1}^\C) = \lfp(T_{\phi_1}^\C) \setminus \lfp(T_{\phi_2}^\C).
    \]

    The first equality follows from \cref{alg:disunif}, \cref{ln:cia-id}.
    The second follows because the heads of $\phi_1$ and $\phi_2$ do not unify.
    As they do not unify, no atom in $\lfp(T_{\phi_2}^\C)$ can be contained in $\lfp(T_{\phi_1}^\C)$.

    Now presume that the heads of $\phi_1$ and $\phi_2$ do unify and let $\theta = \MGU(h_1, h_2)$.
    We proceed to prove \cref{thm:sound-complete-cia} by contradiction.
    Suppose that $\lfp(T_{\phi_1 \setminus \phi_2}^\C) \neq \lfp(T_{\phi_1}^\C) \setminus \lfp(T_{\phi_2}^\C)$.
    By \cref{lem:fact-consequences}, we can rewrite this inequality to:
    \[
        T_{\phi_1 \setminus \phi_2}^\C(\emptyset) \neq T_{\phi_1}^\C(\emptyset) \setminus T_{\phi_2}^\C(\emptyset).
    \]

    Thus, there must exist a ground atom $a$ such that:
    \[
        a \in T_{\phi_1 \setminus \phi_2}^\C(\emptyset) \iff a \notin T_{\phi_1}^\C(\emptyset) \setminus T_{\phi_2}^\C(\emptyset).
    \]

    We proceed by case distinction.
    \begin{itemize}
        \item Presume that $a \in T_{\phi_1 \setminus \phi_2}^\C(\emptyset)$.
        Then, by the above statement, we have that either $a \notin T_{\phi_1}^\C(\emptyset)$ or $a \in T_{\phi_2}^\C(\emptyset)$.
        \begin{itemize}
            \item Presume that $a \notin T_{\phi_1}^\C(\emptyset)$.
            This is impossible by \cref{lem:disunif-in-left}.
            Contradiction.
            \item Presume that $a \in T_{\phi_2}^\C(\emptyset)$.
            This is impossible by \cref{lem:disunif-dif-right}.
            Contradiction.
        \end{itemize}
        \item Now presume that $a \in T_{\phi_1}^\C(\emptyset) \setminus T_{\phi_2}^\C(\emptyset)$.
        Then, symmetrically to the case before, we must have that $a \notin T_{\phi_1 \setminus \phi_2}^\C(\emptyset)$.
        \begin{itemize}
            \item Presume that $a$ and $h_2$ unify.
            Then there must exist a valuation $\sigma$ such that $a = \theta\sigma(h_1) = \theta\sigma(h_2)$ by \cref{lem:extend-mgu}.

            Since $a \notin T_{\phi_2}^\C(\emptyset)$, there must be a constraint $c \in \Psi_2$ such that $D_\C \not\models \theta\sigma(c)$ or equivalently $D_\C \models \theta\sigma(\neg c)$.
            Also observe that since $a \in T_{\phi_1}^\C(\emptyset)$, we have $D_\C \models~\theta\sigma(\Psi_1).$
            Thus:
            \[
                D_\C \models~\theta\sigma(\Psi_1 \cup \{ \neg c \}).
            \]

            Then, for $\psi = \theta(h_1) \lrule \theta(\Psi_1 \cup \{ \neg c \})$, we have that $a \in T_\psi^\C(\emptyset)$ under the valuation $\sigma$.
            The formula $\psi$, though, is in $\phi_1 \setminus \phi_2$ (see \cref{alg:disunif} \cref{ln:cia-unify}), and thus $a \in T_{\phi_1 \setminus \phi_2}^\C(\emptyset)$.
            Contradiction.

            \item Now suppose that $a$ and $h_2$ do not unify.
            Let $\sigma$ be a valuation such that $a = \sigma(h_1)$ and $D_\C \models \sigma(\Psi_1)$.
            As $a$ and $h_2$ do not unify, observe that for all valuations $\sigma'$ we have $a \neq \sigma'(h_2)$.
            Hence $D_\C \models \forall \varsf(h_2). a \neq h_2$.
            Thus, $\sigma$ is a valuation modeling $\psi = h_1 \lrule \{ \forall \varsf(h_2). h_1 \neq h_2 \} \cup \Psi_1$.
            The formula $\psi$, though, is included in $\phi_1 \setminus \phi_2$ (see \cref{alg:disunif} \cref{ln:cia-no-unif}), and therefore $a \in T_{\phi_1 \setminus \phi_2}^\C(\emptyset)$.
            Contradiction.
        \end{itemize}
    \end{itemize}
\end{proof}

\section{Simplifying Inequalities}
\label[appendix]{apx:ineqs}
Our constraint solver implementation simplifies inequality constraints as follows.
It discards inequalities that are true, and it reduces inequalities on two of the same functors to a disjunction of inequalities between corresponding arguments.
For example, it would reduce $f(a, X, Y) \neq f(a, Z, Z)$ to the disjunction $X \neq Z \lor Y \neq Z$ as when either inequality holds, the original inequality holds as well.
Our constraint solver thereby maintains the invariant that inequalities always have a variable on their left-hand side.

Our constraint solver also supports simplifying universally quantified inequality constraints, which result from our change-impact analysis (see \cref{sec:cia}).
Observe that \cref{alg:disunif} only emits universally quantified constraints of the shape $\forall \varsf(t_r). t_l \neq t_r$, i.e., all variables on the inequality's left-hand side are free, and all variables on the inequality's right-hand side are quantified.
Our solver simplifies these constraints in four cases and the constraint remains unchanged in any other case:
\begin{itemize}
    \item If $t_l$ is ground, the constraint is solved by unification.
    \item If $t_r$ is a variable, the constraint evaluates to $\bot.$
    \item If $t_l$ and $t_r$ do not unify, the constraint evaluates to $\top$.
    \item If $t_l$ and $t_r$ use the same function as the top-most symbol, we simplify their inequality to a disjunction of inequalities of their arguments.
    We then distribute the universal quantifier over this disjunction as much as possible.
    Afterwards, we recursively simplify each disjunct.
    For example, our solver simplifies the following inequality as follows:
    \begin{gather*}
        \forall X, Y. f(a, b, A, B) \neq f(a, X, Y, Y) \\
        \text{becomes}~a \neq a \lor \forall X. b \neq X \lor \forall Y. (A \neq Y \lor B \neq Y) \\
        \text{becomes}~\forall Y. (A \neq Y \lor B \neq Y).
    \end{gather*}
\end{itemize}

\section{Well-Formedness Checks}
\label[appendix]{apx:wellformedness}
DecDNS checks that zone files are well-formed before compiling them.
Note that we could enforce stricter well-formedness requirements.
For example, RFC 5155, Sec.~7.2, \cite{NSEC3} requires that \NSECIII records in a given response must share all relevant parameters.
We, though, decided to only enforce invariants that are required for the correctness of our formal model.

\begin{itemize}
    \item A zone must contain exactly one \SOA{} record, which must occur at the apex.
    \item A zone must contain at least one \NS{} at the zone apex.
    \item A zone must not contain \DS{} records at the apex.
    \item If an owner name has a \CNAME{} record, only \CNAME{}, \NSEC{}, and \RRSIG{} records may occur at that owner name \cite[Sec.~3.6.2]{DNS1}.
    \item If a non-apex owner name has an \NS{} record, only \NS{}, \DS{}, \NSEC{}, \RRSIG{}, \A{}, and \AAAA{} records may occur at that owner name, and only \A{} and \AAAA{} records may occur below it.
    \item No owner name may have more than one \CNAME{}, \DNAME{}, or \NSEC{} record.
    \item If an owner name has a \DNAME{} record, no records may occur below it \cite[Sec.~2.4]{DNAMERecord}.
    \item \DNAME{}, \DS{}, and \NS{} records cannot have a wildcard owner name \cite{DNSWildcards}.
    \item An \RRSIG{} record's signer name must be the owner name or one of its ancestors.
    If the signature covers an \SOA{} record, the signer's name must be equal to the owner name.
    If the signature covers a \DS{} record, the signer's name must be a strict ancestor.
    \item An \RRSIG{} record's label count must be correct with respect to \cite[Sec.~3.1.3]{DNSSECSyntax}.
    \item \DNSKEY{} records may only occur at the apex, except for non-\gls{DNSSEC} keys.
\end{itemize}

DecDNS also generates warnings when records in an RRset have different TTLs (excluding \RRSIG{} records from this check), or when an \RRSIG{} record has a different TTL from any record in the RRset it covers.

\felix{Could make an appendix on differences to \cite{CLP}.
For example, I modify the definition of procedural immediate consequences to avoid introduced BF-derivations.}

\end{document}